\documentclass[%
 reprint,
 amsmath,amssymb,
 aps,
]{revtex4-2}

\usepackage{graphicx}
\usepackage{dcolumn}
\usepackage{bm}

\usepackage{float}
\usepackage{placeins}
\usepackage{xcolor}

\begin{document}

\title{Electron scattering and transport in simple liquid mixtures}

\author{G. J. Boyle$^1$}
 \email{Gregory.Boyle@my.jcu.edu.au}
 \author{N. A. Garland$^2$}%
 \author{R. P. McEachran$^3$}%
\author{K. A. Mirihana$^1$}%
\author{R. E. Robson$^1$}%
\author{J. P. Sullivan$^3$}%
\author{R. D. White$^1$}%
\affiliation{%
$^1$College of Science and Engineering, James Cook University, Townsville, QLD 4814, Australia
}%
\affiliation{%
$^2$Centre for Quantum Dynamics, Griffith University, Nathan, QLD 4111, Australia
}%
\affiliation{%
$^3$Positron Research Group, Australian National University, Canberra, ACT 0200, Australia
}%


\begin{abstract}

The theory for electron transport in simple liquids developed by Cohen and Lekner~\cite{,CohenLekner1967}, is extended to simple liquid mixtures. The focus is on developing benchmark models for binary mixtures of hard-spheres, using the Percus-Yevick model~\cite{Lebowitz1964,Hiroike1969} to represent the density structure effects. A multi-term solution of Boltzmann’s equation is employed to investigate the effect of the binary mixture structure on hard-sphere electron scattering cross-sections and transport properties, including the drift velocity, mean energy, longitudinal and transverse diffusion coefficients. Benchmark calculations are established for electrons driven out of equilibrium by a range of reduced electric field strengths ($0.1-100$ Td). 
\end{abstract}

\maketitle

\section{Introduction\label{sec:Intro}}

An understanding of the behaviour of free electrons in structured fluid systems, such as dense gases and liquids, is of interest from both a fundamental research perspective, and for technological applications.

For example, the 2021 European Committee for Future Accelerators (ECFA) Detector Research and Development Roadmap~\cite{EFCA2021} highlighted the importance of liquid detectors, notably using liquid helium, argon and xenon, for dark matter searches, neutrino physics and astroparticle experiments. The use of dopants and liquid mixtures to improve the scintillation yield, ionisation yield and free electron lifetime is an active area of detector research.

Previous investigations have focused on electron transport in pure noble liquids, including argon~\cite{BoyleEtal2015}, krypton~\cite{WhiteEtal2018} and xenon~\cite{BoyleEtal2016}, based on the framework proposed by Cohen and Lekner in 1967~\cite{CohenLekner1967,Lekner1967}. In this work, we extend the original formalism to dense fluid mixtures. The focus is on binary mixtures, but the extension to arbitrary multi-component mixtures is, in principle, straightforward.

There are important considerations that need to be made in the study of electrons in dense fluid systems that are not significant in dilute gaseous systems. When the de Broglie wavelength of the electrons (near thermal energies) is comparable to the interatomic spacing of the medium, the scattering occurs from multiple neighboring scattering centres simultaneously, which are correlated in space and time. This `coherent scattering' process is in contrast to binary scattering process in dilute gaseous systems, and extension to mixtures is the focus of the current work. A second consideration is that, even within a semi-classical picture where free electrons can be regarded as point particles, in a dense system the electron is never effectively in `free space'. Instead, when scattering from a `target' atom there is always a contribution from the `bulk', requiring modifications to the scattering potential~\cite{Lekner1967}. In this work, we employ short-range analytic hard-sphere models for the electron-atom scattering, and thus long-range potential modifications are not necessary. However, they are critically important when extending the results presented here to real liquid mixture systems, as has been considered elsewhere~\cite{BoyleEtal2015,BoyleEtal2016,WhiteEtal2018}.  

In section~\ref{sec:Struc}, the structure of binary-component liquids is discussed, with a focus on benchmark hard-sphere systems characterized by the Percus-Yevick model. The effect of the relative hard-sphere size and relative number density of the two species components on the partial static structure factors and pair correlation functions is investigated. In section~\ref{sec:Kinetic}, the kinetic theory framework of Cohen and Lekner~\cite{CohenLekner1967}, and generalized to a multi-term framework~\cite{WhitRobs11,Boyl17}, is extended to simple liquid mixtures, and a coherent elastic scattering collision operator that accounts for both the self- and cross-correlation effects is introduced. A multi-term Boltzmann equation solution accounting for anisotropic coherent scattering is employed to calculate transport properties, including the drift velocity, mean energy, transverse and longitudinal diffusion coefficients. These calculations elucidate the effect of the mixture structure parameters on macroscopic electron transport, establishing calculation benchmarks. The inadequacy of Blanc's law~\cite{Blanc1908}, a common approximation used in dilute noble gas applications, for liquid mixtures is also demonstrated.

\section{Structure of simple liquid mixtures\label{sec:Struc}}

\subsection{Theory}

In liquids and dense gases, the atoms/molecules move within a volume subject to correlations between the constituent positions, resulting in some short-range order. In the dilute-phase limit, or at large distances, the motion of the atoms/molecules is uncorrelated. For a binary component mixture, we shall designate the two components with labels `a' and `b', respectively. The structure of a mixture can be characterized by the partial pair distribution functions, $g_{\nu\mu}(\mathbf{r})$, via~\cite{HansenMcDonald1973},
\begin{align}
    x_{\nu}x_{\mu}n_0 g_{\nu\mu}(\mathbf{r}) = \frac{1}{N_{\nu} \! + \! N_{\mu}}\!\left\langle \sum_{i=1}^{N_{\nu}}\sum_{j=1}^{N_{\mu}}\delta(\mathbf{r} + \mathbf{r}_i - \mathbf{r}_j  )\!\right\rangle, \label{eq:g}
\end{align}
where $\mathbf{r}_{i}$ ($\mathbf{r}_{j}$) is the position of the $i$th ($j$th) atom/molecule of type $\nu$ ($\mu$), of which there are $N_\nu$ ($N_\mu$) in total. The total number density of the mixture is $n_0$, and the number density fractions of the two species are given by $x_{\nu} = \frac{N_{\nu}}{N_{\nu}+N_{\mu}}$ and $x_{\mu} = \frac{N_{\mu}}{N_{\nu}+N_{\mu}}$, respectively. Defining the pair correlation function as $h_{\nu\mu}(r) = g_{\nu\mu}(r) - 1$, then the partial pair correlation functions can be related to the partial static structure factors, $S_{\nu\mu}(\mathbf{k})$, via the Fourier transform,
\begin{align}
    S_{\nu\mu}(\mathbf{k}) &= x_{\nu}\delta_{\nu\mu} + x_{\nu}x_{\mu} n_0\!\!\int\!   \exp(-i\mathbf{k}\cdot\mathbf{r})h_{\nu\mu}(\mathbf{r})d\mathbf{r}, \label{eq:S}
\end{align}
where $\mathbf{k}$ is a wavevector. It should be noted that there is an alternative definition of the partial static structure factor, $S^{AL}_{\nu\mu}(\mathbf{k})$, that is prevalent in the literature (e.g. see Ashcroft and Langreth 1967~\cite{AshcroftLangreth1967}), and that the two are related by
\begin{align}
    S_{\nu\mu}(\mathbf{k}) = \sqrt{x_\nu x_\mu}S^{AL}_{\nu\mu}(\mathbf{k}).
\end{align}

If the system is isotropic, $S_{\nu\mu}(\mathbf{k})$ is a function of only the wavenumber $k=|\mathbf{k}|$, and thus
\begin{align}
     S_{\nu\mu}(k) &= x_{\nu}\delta_{\nu\mu} + x_{\nu}x_{\mu}4\pi n_0 \!\! \int_0^{\infty} \!\! h_{\nu\mu}(r)\frac{\sin{(kr)}}{kr}r^2 dr. \label{eq:Siso}
\end{align}

Generally the pair distribution functions and structure factors are generated in computer simulations~\cite{GunsterenBeredsen90,Lindgard1994,WedbergEtal2011}. However, an analytic representation of the static structure factor for hard-sphere liquids can be obtained from the Percus-Yevick equation~\cite{PercusYevick1958}. The exact solution for the  Percus-Yevick hard-sphere equation for binary mixtures was found by Lebowitz, and for multi-component mixtures by Lebowitz~\cite{Lebowitz1964} and Hiroike ~\cite{Hiroike1969}.

The structure of a binary hard-sphere Percus-Yevick liquid mixture is completely specified
by the hard-sphere diameter $d_a$ of species $a$, the density fraction $x_a$ of species $a$, the ratio $\beta = d_b/d_a$, where $d_b$ is the  hard-sphere diameter of species $b$, along with the total packing factor
\begin{align}
    \Phi = \frac{\pi}{6}n_0 x_ad_a^2 + \frac{\pi}{6} n_0 x_bd_b^2. \label{eq:phi}
\end{align}
 Note that $x_b = 1 - x_a$, and $n_0$ is the total number density of the binary mixture.
This convention is adopted for the remainder of this paper, where the hard-sphere Percus-Yevick model is used to represent the partial static structure factors, with the aim to investigate the effect of the structure properties on electron-atom scattering and transport. 

\begin{figure*}[ht]
\centering
\centering
    \includegraphics[width=1.0\linewidth, angle=0]{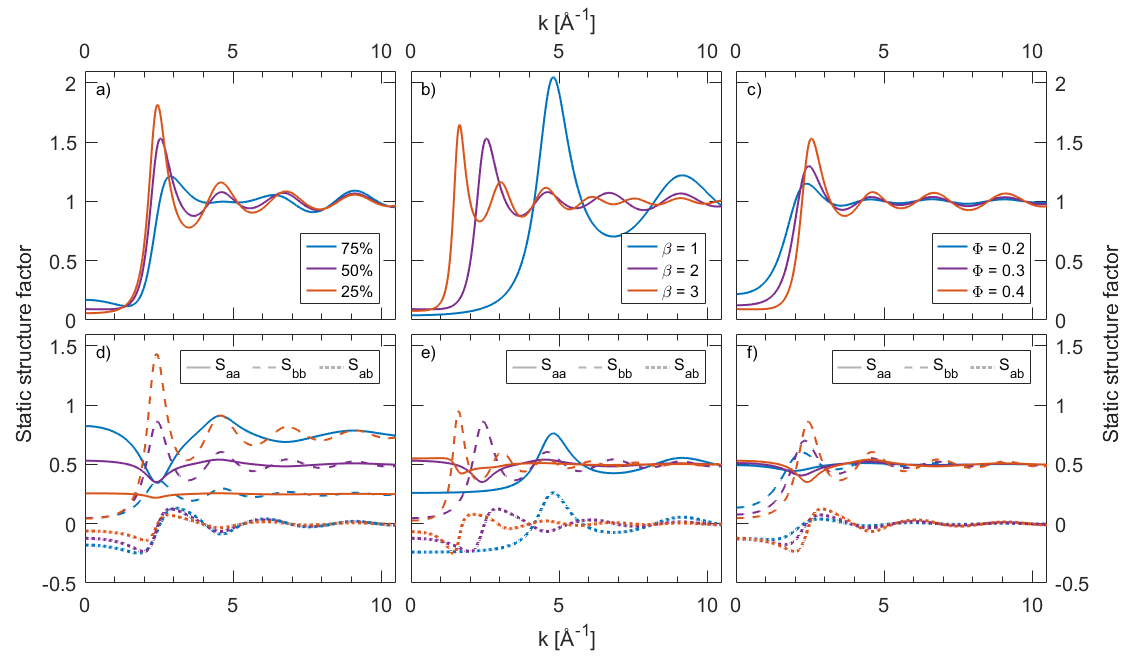}
    \caption{Structure properties of binary hard-sphere mixtures. Panels (a)-(c) show the total static structure factor as defined by equation~\eqref{eq:S_tot}. Panels (d)-(f) show the partial static structure factors as defined by equation~\eqref{eq:Siso}. \textbf{a)} Total static structure factor for varying percentages of species $a$, for $\Phi = 0.4$ and $\beta = 2$. \textbf{b)} Total static structure factor for varying values of $\beta$, for $\Phi = 0.4$ and $x_a = 0.5$. \textbf{c)} Total static structure factor for varying packing factors, for $\beta=2$, and $x_a = 0.5$. \textbf{d)} Partial static structure factors for varying percentages of species $a$, for $\Phi = 0.4$ and $\beta = 2$. \textbf{e)} Partial static structure factors for varying values of $\beta$, for $\Phi = 0.4$ and $x_a = 0.5$. \textbf{f)} Partial static structure factors for varying packing factors, for $\beta=2$, and $x_a = 0.5$. 
    } 
    \label{fig1}
\end{figure*} 

\subsection{Benchmark system structure factors}

 We now consider a two-component, hard-sphere mixture, with the hard-sphere diameter of species $a$ set to $d_a = \sqrt{\frac{6}{\pi}}$ (for consistency with the electron scattering model introduced in Section~\ref{sec:Kinetic}), and the parameters $\Phi$, $x_a$ (with $x_b = 1-x_a$), and $\beta = d_b/d_a$ allowed to vary. For hard-sphere scattering, where the magnitude of the scattering cross-section is proportional to the square of the hard-sphere radius, we can define a total static structure factor, $S_{tot}(k)$, defined by~\cite{Hoshino1983},
\begin{align}
    S_{tot}(k) = \frac{S_{aa}(k) + 2\beta S_{ab}(k) + \beta^2 S_{bb}(k)}{x_a + \beta^2x_b}. \label{eq:S_tot}
\end{align}
The variation of the total static structure factor for various parameter sets is shown in Figures~\ref{fig1}\,(a)-(c), while variation of the partial structure factor components defined by Equation~\eqref{eq:Siso} is shown Figures~\ref{fig1}\,(d)-(f). The total static structure factors, $S_{tot}$, are shown to oscillate around unity, while the partial static structure factors, $S_{\nu\mu}$, oscillate around $x_a$, $x_b$ and $0$ for the $S_{aa}$, $S_{bb}$ and $S_{ab}$ components respectively, and approach these values in the limit of large $k$, as required. 

In Figures~\ref{fig1}\,(a) and (d), the total packing factor is set to $\Phi = 0.4$, the hard-sphere ratio is set to $\beta = 2$, while the number density fraction varies as $x_a = 0.75,\, 0.5,\, 0.25$. Species $b$ has a relatively larger hard-sphere radius, and as the fraction of a decreases (and the fraction of $b$ increases), the total static structure of the mixture tends towards larger amplitudes and shifts towards lower values of $k$. This can also be seen in the partial static structure factors, where the component $S_{bb}$ demonstrates sharper oscillations than the component $S_{aa}$. The slight uptick in the value of the total static structure factor at low $k$ values for $x_a = 0.75$ is a manifestation of the large amplitude values of $S_{aa}$. The connection between the pair correlation function and static structure factor by the Fourier transform in equation~\eqref{eq:Siso} means that the location of the first peak in the static structure factor is shifted to lower values of $k$ as the hard-sphere diameter increases.

In Figures~\ref{fig1}\,(b) and (e), the total packing factor is set to $\Phi = 0.4$, the number density fraction is set to $x_a = 0.5$, while the hard-sphere diameter varies as $\beta = 1, 2, 3$. For $\beta = 1$, the hard-sphere diameters of each species is the same, and thus represents a pure liquid. As the value of $\beta$ increases, the peak of each of the static structure factor components shift to lower $k$ values, and the wavelength of the oscillations reduces considerably.

In Figures~\ref{fig1}\,(c) and (f), the hard-sphere ratio is set to $\beta = 2$, the number density fraction is set to $x_a = 0.5$, while the total packing factor varies as $\Phi = 0.2,\, 0.3,\, 0.4$. As the value of $\Phi$ increases and the liquid mixture increases its density, the amplitude of each of the static structure factor components increases, showing large differences from the dilute case of unity.

The partial static structure factor directly plays a pivotal role in describing coherent scattering effects, and the kinetic theory and transport of such mixtures is detailed in the following section. 

\section{Coherent scattering in simple mixtures\label{sec:Kinetic}}

\subsection{Kinetic theory}

The behaviour of electrons in a gaseous or liquid system, driven out of equilibrium via an electric field $\mathbf{E}$, can be described by the solution of the Boltzmann’s equation for the phase-space distribution function $f$:
\begin{align}
    \frac{\partial f}{\partial t} + \bm{v}\cdot\nabla f + \frac{e\bm{E}}{m_e}\cdot\frac{\partial f}{\partial \bm{v}} &= - J(f), \label{eq:Boltz}
\end{align}
where $\bm{r}$, $\bm{v}$, and $e$ denote the position, velocity, and charge of the electron, respectively. The collision operator $J(f)$ accounts for interactions between the electrons of mass $m_e$ and the background material. 

If there are no strong gradients, then the space-time dependence of the phase-space distribution function can be projected onto the number density $n(\bm{r},t)$ via the density gradient expansion~\cite{KumaSkulRobs80},
\begin{align}
    f(\bm{r},\bm{v},t) &= nF(\bm{v}) - F^{L}(\bm{v})\frac{\partial n}{\partial z} \nonumber \\ 
     & \ \ \ \ - F^{T}(\bm{v})\left[\cos\phi \frac{\partial n}{\partial x} + \sin\phi \frac{\partial n}{\partial y} \right]  + \dots, \label{eq:DensGradExp}
\end{align}
where $\phi$ is the azimuthal angle, and the superscripts $L$ and $T$ define quantities that are parallel and transverse to the electric field, respectively. 

Solution of Boltzmann’s equation~\eqref{eq:Boltz} with the expansion~\eqref{eq:DensGradExp} requires decomposition of the coefficients in velocity space through an expansion in (associated) Legendre polynomials, $P_l^m(\cos\theta)$~\cite{AbramowitzStegun1972}, i.e.,
\begin{align}
    F(\bm{v}) &= \sum_{l=0}^\infty F_l(v)P_l(\cos\theta), \\
    F^{L}(\bm{v}) &= \sum_{l=0}^\infty F_l^{L}(v)P_l^0(\cos\theta), \\
    F^{T}(\bm{v}) & = \sum_{l=0}^\infty F_l^{T}(v)P_l^1(\cos\theta),
\end{align}
where $\theta$ denotes the angle relative to the electric field direction. 

Boltzmann's equation can then be re-written as a hierarchy of equations for these expansion coefficients:
\begin{widetext}
\begin{align}
   J_l(F_l) \! + \! \frac{l\!+\!1}{2l\!+\!3}\!\left(\frac{qE}{m_e}\right)\!\left(\frac{\partial}{\partial v} \! + \! \frac{l\!+\!2}{v}\right)\!F_{l+1} \!
    + \! \frac{l}{2l\!-\!1}\!\left(\frac{qE}{m_e}\right)\!\left(\frac{\partial}{\partial v} \!-\! \frac{l\!-\!1}{v}\right)\!F_{l-1} &= \! 0, \label{eq:F} \\ 
    J_l(F_l^{T}) \! + \! \frac{l\!+\!2}{2l\!+\!3}\!\left(\frac{qE}{m_e}\right)\!\left(\frac{\partial}{\partial v} \! + \! \frac{l\!+\!2}{v}\right)\!F_{l+1}^{T} \!
    + \! \frac{l-1}{2l\!-\!1}\!\left(\frac{qE}{m_e}\right)\!\left(\frac{\partial}{\partial v} \!-\! \frac{l\!-\!1}{v}\right)\!F_{l-1}^{T}  &= v\!\left(\frac{1}{2l-1}F_{l-1} \! - \! \frac{1}{2l\!+\!3}F_{l+1}\right), \label{eq:FT} \\
    J_l(F_l^{L}) \! + \! \frac{l\!+\!1}{2l\!+\!3}\!\left(\frac{qE}{m_e}\right)\!\left(\frac{\partial}{\partial v} \! + \! \frac{l\!+\!2}{v}\right)\!F_{l+1}^{L} \!
    + \! \frac{l}{2l\!-\!1}\!\left(\frac{qE}{m_e}\right)\!\left(\frac{\partial}{\partial v} \!-\! \frac{l\!-\!1}{v}\right)\!F_{l-1}^{L}  &= v\!\left(\frac{l+1}{2l+3}F_{l+1} \! + \! \frac{l}{2l\!-\!1}F_{l-1}\right) \! - \! WF_l, \label{eq:FL} 
\end{align} 

\end{widetext}
where the $J_l$ represent the Legendre projections of the collision operator and $W$ is the flux drift velocity defined in~\eqref{eq:W}.

Solving for the (normalised) coefficients of equations~\eqref{eq:F}--\eqref{eq:FL} provides sufficient information to calculate transport properties, such as the (flux) drift velocity $W$, mean energy $\epsilon$, (flux) longitudinal diffusion coefficient $D_L$ and (flux) transverse diffusion coefficient $D_T$: 
\begin{align}
    W = \frac{4\pi}{3}\int_0^\infty v^3F_1 dv, \label{eq:W}
\end{align}
\begin{align}
    \epsilon =2\pi m_e\int_0^\infty v^4F_0 dv, \label{eq:En}
\end{align} 
\begin{align}
    D_{L,T} = \frac{4\pi}{3}\int_0^\infty v^3F_1^{L,T} dv. \label{eq:DLT}
\end{align}
In this work, there are no non-conservative scattering processes operative, and thus flux and bulk properties are identical~\cite{RobsWhitHild17}.

We shall henceforth consider only low-energy scattering from atomic systems such that only elastic collisions are operative. For dense systems, the de Broglie wavelength of the electron can be comparable to or smaller than the inter-particle spacing of the medium, $~N^{-1/3}$. As such, the low-energy electron can be viewed as a wave that simultaneously interacts with multiple scattering centres in the medium. At higher energies, the de Broglie wavelength decreases and this coherent scattering effect diminishes until the binary scattering approximation is recovered. For liquid helium, the average interparticle spacing is $\approx 2.5\, \AA$, implying that `low' energies are those less than $\approx 0.6$ eV.

In order to account for the low-energy coherent scattering, the elastic collision operator must be modified~\cite{CohenLekner1967, RobsWhitHild17}. In Appendix~\ref{sec:DCS} we derive the differential cross-sections for liquid mixtures with coherent scattering effects, as well as discuss the corresponding generalization to the elastic collision operator. The Legendre projections of the elastic collision operator for electrons, accounting for coherent scattering, are given for a binary mixture by:
\begin{align}
    J_0 (\Phi_0) &= \frac{m_e}{m_a}\frac{1}{v^2}\left[v\nu^{a}_1(v)\left(v\Phi_l + \frac{k_bT}{m_e}\frac{d}{dv}\Phi_0\right)\right] \nonumber \\
    &\ + \frac{m_e}{m_b}\frac{1}{v^2}\left[v\nu^{b}_1(v)\left(v\Phi_l + \frac{k_bT}{m_e}\frac{d}{dv}\Phi_l\right)\right], \label{eq:J0} \\
    J_l (\Phi_l) &= \tilde{\nu}_l(v) \Phi_l,\ \ \ \ \text{for } l \geq 1, \label{eq:Jl} 
\end{align} 
where $\Phi_l= {F_l,\,F_l^{T},F_l^{L}}$, and $m_a$ and $m_b$ are the masses of species $a$ and $b$, respectively, while
\begin{align}
    \frac{\nu^{a,b}_l(v)}{x_{a,b}n_0} &= 2\pi v\!\!\int_0^{2\pi}\!\!\sigma_{a,b}(v,\theta)[1-P_l(\cos\theta)]\sin\theta d\theta,
\end{align}
are the usual binary collision frequencies for each species individually, while
\begin{align}
    \frac{\tilde{\nu}_l(v)}{n_0} &= 2\pi v\!\!\int_0^{2\pi}\!\!\Sigma(v,\theta)[1-P_l(\cos\theta)]\sin\theta d\theta,
\end{align}
is the total structure-modified collision frequency, corresponding to the total structure modified differential cross-section (see Appendix~\ref{sec:ColOpApp}),
\begin{align}
    \Sigma(v,\theta)
    &=  \sigma_a(v,\theta)S_{aa}\left(\frac{2m_ev}{\hbar}\sin{\frac{\theta}{2}}\right)   \nonumber \\ 
    &\ \ \ + \sigma_b(v,\theta) S_{bb}\left(\frac{2m_ev}{\hbar}\sin{\frac{\theta}{2}}\right) \nonumber \\
    &\ \ \ + 2 \sqrt{\sigma_a(v,\theta)\sigma_b(v,\theta)} S_{ba}\left(\frac{2m_ev}{\hbar}\sin{\frac{\theta}{2}}\right). \label{eq:DCS}
\end{align}
In the limit of large argument values, $S_{aa} \rightarrow x_a, S_{bb} \rightarrow x_b$ and $S_{ab} \rightarrow 0$, such that the usual dilute gas mixture case is recovered, as required. In the case of $b \rightarrow a$, $\sigma_b \rightarrow \sigma_a$ and $S_{aa} + 2S_{ab} + S_{bb} \rightarrow S_a$, where $S_a$ represents the structure factor for a pure liquid of species $a$, as required.

Finally, we introduce the partial cross-sections $\sigma_l(v)$ and $\Sigma_l(v)$ defined by~\cite{RobsWhitHild17},
\begin{align}
    \sigma_l(v) &= 2\pi\int_{-1}^1 \sigma(v,\theta)P_l(\cos\theta) d(\cos\theta), \label{eq:sig_l} \\
    \Sigma_l(v) &= 2\pi\int_{-1}^1 \Sigma(v,\theta)P_l(\cos\theta) d(\cos\theta), \label{eq:Sig_l}
\end{align}
such that the collision frequencies in equations~\eqref{eq:J0}--\eqref{eq:Jl} can be represented by,
\begin{align}
    \nu_l(v) &= n_0v[\sigma_0(v) - \sigma_l(v)], \label{eq:nu_l}
\end{align}
and
\begin{align}
    \tilde{\nu}_l(v) &= n_0v[\Sigma_0(v) - \Sigma_l(v)]. \label{eq:Nu_l}
\end{align}

Having defined the connection between the microscopic scattering cross-sections and macroscopic transport properties, we are now in a position to investigate the influence of structure parameters on the electron transport. 

\subsection{Benchmark system electron scattering and transport \label{sec:Benchmark}}

\textbf{
\begin{figure}[b]
\centering
\centering
    \includegraphics[width=1.0\linewidth, angle=0]{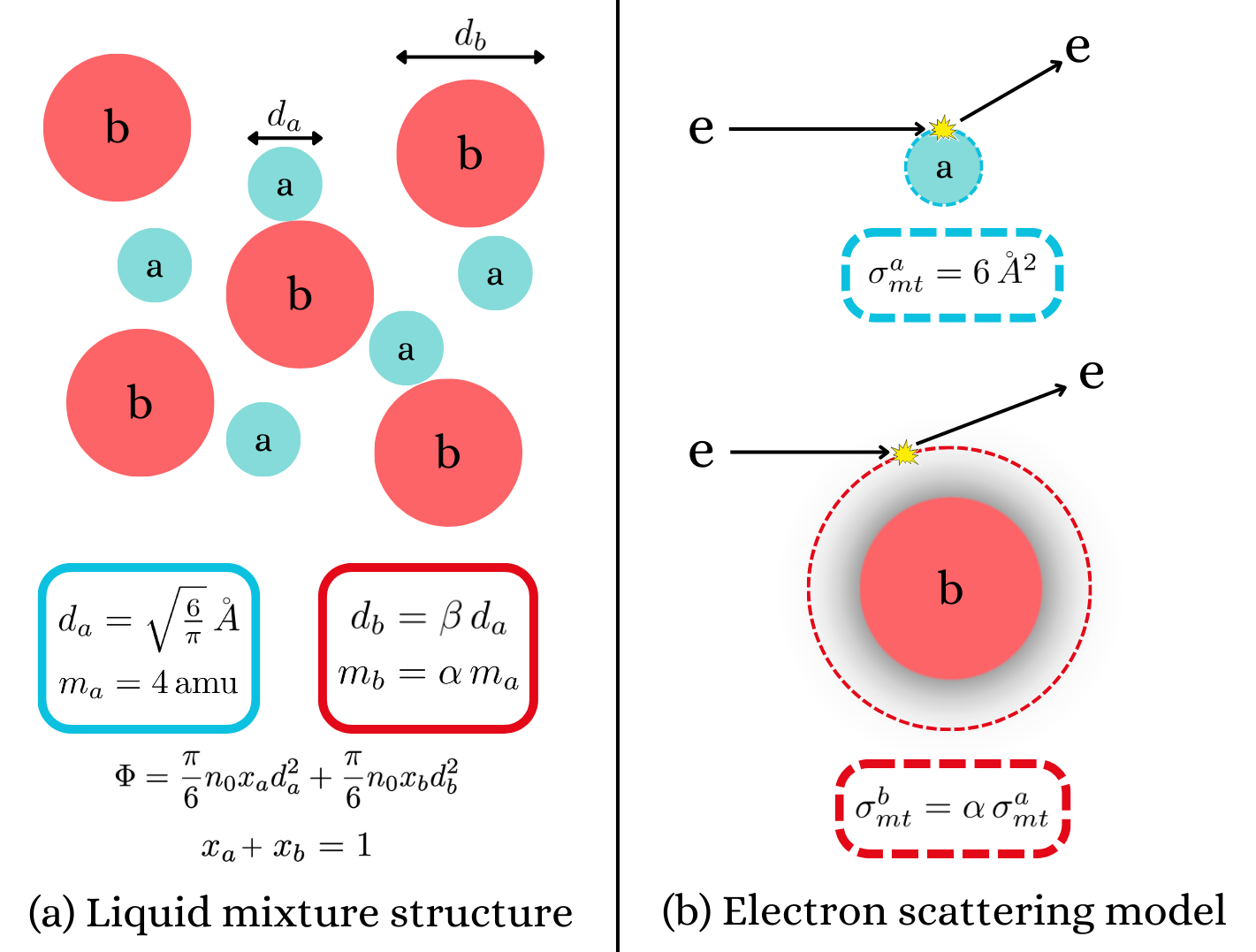}
    \caption{Benchmark binary system of cold ($T=0$ K) hard-spheres. \textbf{a)} Parameters describing the liquid mixture structure: $d_a$ ($d_b$), $m_a$ ($m_b$) and $x_a$ ($x_a$) are the hard-sphere diameter, mass and density fraction of species $a$ (species $b$), respectively. $\Phi$ is the total packing factor, $n_0$ is the total number density. The values of $\alpha$ and $\beta$ are varied to alter the structure properties of species $b$. \textbf{b)} Parameters describing the electron scattering: $\sigma_{mt}^a$ ($\sigma_{mt}^b$) is the momentum-transfer cross-section for electron scattering from species $a$ (species $b$), and $\sigma_0^a-\sigma_l^a = \sigma_{mt}^a$ ($\sigma_0^b-\sigma_l^b = \sigma_{mt}^b$) is constant for all $l$. The value of $\alpha$ is varied to alter the electron scattering properties of species $b$. Note that, despite the simple representation here, the electron interaction is a wave scattering process.}
    \label{fig0}
\end{figure} }

In this section, we will consider a benchmark cold \mbox{($T=0$)} binary system of hard-spheres, as detailed in Figure~\ref{fig0}. The values of $\alpha$ and $\beta$, along with the total packing factor, $\Phi$, and the density fraction, $x_a$ (with $x_b = 1 - x_a$), are parameters to be varied. The parameters $\alpha$ and $\beta$ relate to the size of the background atoms with respect to electron scattering and atomic scattering, respectively. The set value of $d_a$ has been chosen to correspond with $\sigma_{mt}^a$ for hard-sphere scattering ($\sigma_{mt}=\pi d^2$~\cite{Fitzpatrick2015}), and under this picture of electrons scattering with a liquid mixture of hard-spheres, $\alpha = \beta^2$ for consistency. In general, we will allow $\alpha$ and $\beta$ to vary independently to highlight the influence of each parameter. The constraint that $m_b = \alpha\,m_a$ has the property of keeping the elastic energy exchange collision frequency, $ \sim\!\frac{2m}{m_0}\nu_{1}$, constant as we vary the cross-section of species $b$. When $x_a = 1$, the benchmark system above reduces to the pure liquid system investigated previously~\cite{WhitRobs11,TattersallEtal2015,Boyl17}.

\begin{figure}[t]
\centering
\centering
    \includegraphics[width=1.0\linewidth, angle=0]{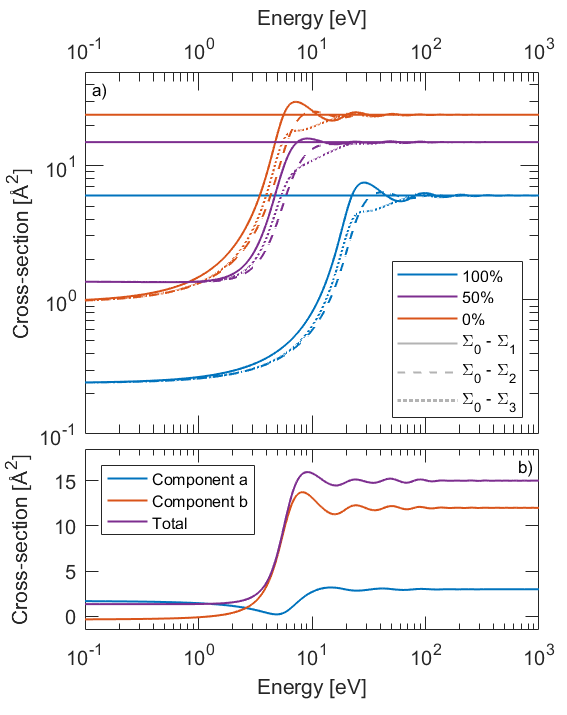}
    \caption{Scattering cross-section variation with electron energy for the binary hard-sphere mixture given in Figure~\ref{fig0}.  \textbf{a)} Partial cross-sections $\Sigma_0 - \Sigma_l$, as defined by equation~\eqref{eq:Sig_l}, for $l = 1,2,3$, for varying percentages of species $a$, and with parameters $\Phi = 0.4$, $\alpha = \frac{\sigma_b}{\sigma_a} = 4$, $\beta = \frac{d_b}{d_a} = 2$. Horizontal lines represent the (constant) binary-scattering limit cross-sections. \textbf{b)} Momentum-transfer cross-section components as defined by equation~\eqref{eq:DCS_separate}, for $\Phi = 0.4$, $\alpha = \frac{\sigma_b}{\sigma_a} = 4$, $\beta = \frac{d_b}{d_a} = 2$ and $x_a = 50\%$.} 
    \label{fig2}
\end{figure} 

\begin{figure}[h!]
\centering
\centering
    \includegraphics[width=1.0\linewidth, angle=0]{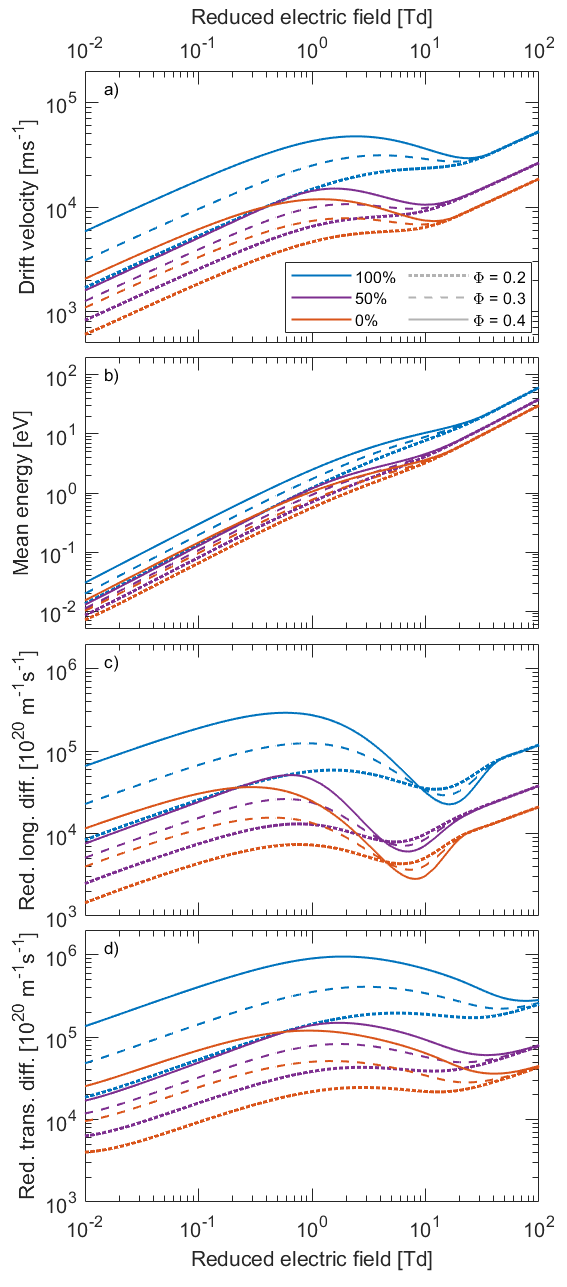}
    \caption{Transport properties (defined in equations~\eqref{eq:W}--\eqref{eq:DLT}) as a function of reduced electric field, $E/n_0$, with varying  mixture percentage of species $a$. The parameters of species $a$ and $b$ are set according to the model given in Figure~\ref{fig0}, with $\alpha = \frac{\sigma_b}{\sigma_a}=4$ and $\beta = \frac{d_b}{d_a}=2$. \textbf{a)} Drift velocity, \\ \textbf{b)} Mean energy, \textbf{c)} Reduced longitudinal diffusion coefficient, \textbf{d)} Reduced transverse diffusion coefficient.}
    \label{fig3}
\end{figure} 

In Figure~\ref{fig2}\,(a), the constant total binary cross-section $\sigma_0 - \sigma_1 = (x_a + x_b\alpha)6\, \AA^2$, along with the structure modified components $\Sigma_0 - \Sigma_l$, for $x_a = 1, 0.5$ and $0$, are presented. As a reflection of the static structure factor behaviour, the profiles oscillate around the constant binary-scattering value. As the fraction of species $b$ increases, which has a larger cross-section magnitude, so does the mixture cross-section magnitude. The high-order partial terms demonstrate similar behaviour, with the oscillations appearing to dampen as the order increases. As in the pure liquid case, the key feature is the greatly-reduced momentum transfer at low energies, which we expect to result in enhanced transport properties such as drift and diffusion at low electric field strengths.

It may seem natural to re-write the effective cross-section in equation~\eqref{eq:DCS} as,
\begin{align}
    \Sigma(v,\theta) &= \ \sigma_a(v,\theta)\left[S_{aa} + \sqrt{\frac{\sigma_b(v,\theta)}{\sigma_a(v,\theta)}} S_{ba} \right] \nonumber \\ 
    &\ \ + \sigma_b(v,\theta)\left[S_{bb} + \sqrt{\frac{\sigma_a(v,\theta)}{\sigma_b(v,\theta)}} S_{ba}  \right],\label{eq:DCS_separate} 
\end{align} 
which may then be considered as the `effective' scattering from species $a$ and from species $b$ individually, while also accounting for the influence of neighboring atoms. However, even though positivity of the total combined cross-section is ensured, the positivity of the individual components can be violated, which challenges this intuitive splitting. The components defined in equation~\eqref{eq:DCS_separate} are shown in Figure~\ref{fig2}\,(b), and it can be seen that the cross-section component belonging to species $b$ is negative for energies below $\approx 1$\, eV. Note that is not a problem as the total cross-section for the mixture is indeed always positive.

\begin{figure*}[ht]
\centering
\centering
    \includegraphics[width=1.0\linewidth, angle=0]{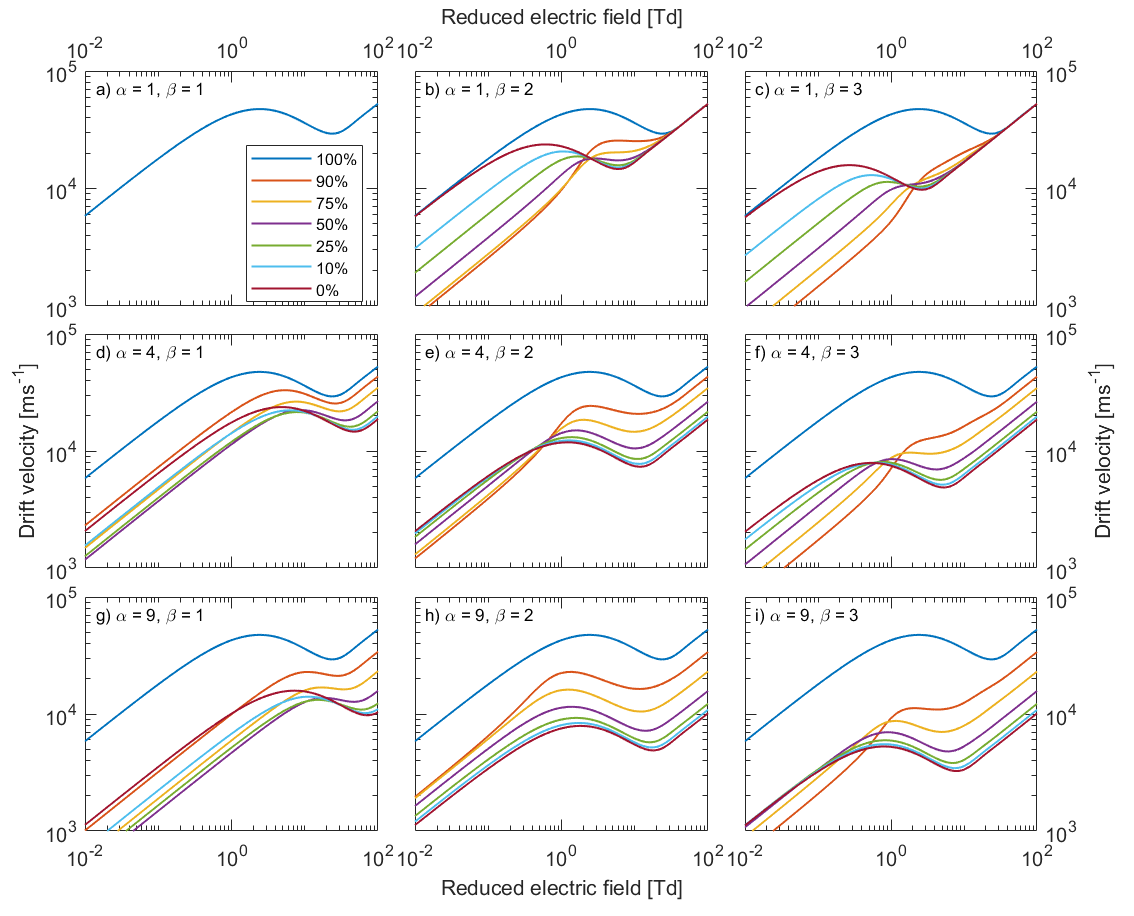}
    \caption{Drift velocity (defined in equation~\eqref{eq:W}) as a function of reduced electric field, $E/n_0$, with varying  mixture percentage of species $a$, for $\Phi=0.4$. The parameters of species $a$ and $b$ are set according to the model given in Figure~\ref{fig0}, with the values of $\alpha = \frac{\sigma_b}{\sigma_a}$ and $\beta = \frac{d_b}{d_a}$ stated in each panel. The rows represent increasing $\alpha$ moving from top to bottom. The columns represent increasing $\beta$ moving from left to right. Note that when $\alpha = 1$ and $\beta = 1$, the drift velocity is the same for each mixture percentage, and hence only a single profile is shown in panel (a).}  
    \label{fig4}
\end{figure*} 

The variation of transport coefficients ($W$, $\epsilon$, $n_0D_L$, $n_0D_T$) with reduced electric field strength for $\beta = 2$ and $\alpha = \beta^2 = 4$, for a selection of packing factors $\Phi$ and density fractions $x_a$ is shown in Figure~\ref{fig3}. Orders of magnitude differences in the transport properties is evident between the various mixture systems, highlighting the importance of an accurate description of the structure, scattering and transport. It is clear that the oscillatory nature of the structure factors, and subsequently the structure-modified cross-sections is reflected in the behaviour of the transport properties. Similarly, it is clear that at high reduced electric field strengths, and thus high mean energies, the structure effects are suppressed and the profiles corresponding to different packing factors converge. The relatively lower momentum-transfer at low-energies for collisions involving species $a$ compared to those involving species $b$ results in a greater enhancement of the transport properties corresponding to larger values of $x_a$. Increasing the packing factor also tends to decrease the momentum-transfer, and once again a greater enhancement of the transport properties is seen for larger packing factors compare to smaller packing factors. For $\Phi = 0.4$, the drift velocity profiles in Figure~\ref{fig3}\,(a) demonstrate the phenomenon of structure-induced negative differential conductivity (NDC)~\cite{WhiteRobs09} between $\approx 2.5-25$ Td, i.e., a \textit{decrease} in the drift velocity despite an \textit{increase} in the reduced electric field strength. The NDC is reduced or eliminated entirely for the smaller volume fractions.

The drift velocity as a function of electric field strength for $\Phi = 0.4$, and select density fractions $x_a$, parameters $\alpha$ and $\beta$ are shown in Figure~\ref{fig4}. The rows represent increasing $\alpha$ values, while the columns represent increasing $\beta$ values. When both $\alpha = 1$ and $\beta = 1$, the system represents a pure liquid, and there is no variation with the density fractions, as demonstrated by the single profile in Figure~\ref{fig4}\,(a).

The explicit effect of the mixture structure, rather than the electron scattering magnitude, is isolated in Figures~\ref{fig4}\,(b) and (c), where $\alpha = 1$ and $\beta > 1$. Generally, increasing $\beta$ has the effect of decreasing the drift velocity for a given density fraction. Complicated behaviour is demonstrated by the profiles for the various density fractions. When $x_a = 0.9$, there is already a very significant drop in the drift velocity at low $E/n_0$. As $x_a$ increases, the drift velocity then \textit{rises} at these low fields, until $x_a = 0$ reaches the same drift velocity as the $x_a = 1$ case. At moderate fields, the oscillatory nature of the profiles varies in magnitude, with there being an apparent intersection point between the profiles for $x_a \neq 1$ at approximately $E/n_0 = 2$ Td in both figures. At high fields $\gtrapprox 30$ Td, the profiles converge to the pure binary-scattering case. In Figure~\ref{fig4}\,(c), NDC is absent for $x_a = 0.9-0.5$, and re-emerges for $x_a \leq 0.25$.

The explicit effect of the electron scattering cross-section magnitude ratio, rather than the relative hard-sphere size of the background, is isolated in Figures~\ref{fig4}\,(d) and (g), where $\beta = 1$ and $\alpha > 1$. At high $E/n_0 \gtrapprox 50$ Td, decreasing $x_a$ decreases the magnitude of the drift velocity monotonically, and this effect is enhanced for larger $\alpha$. This is consistent with the binary-scattering picture, where increasing the total momentum-transfer of the mixture by increasing either $x_b$ or $\alpha$ acts to decrease the corresponding drift velocity. At lower reduced electric field strengths, the behaviour is once again complicated by the structure influence, and does not display monotonic trends. There is a clear decrease in the drift velocity for $x_a < 1$ when compared to the pure $x_a = 1$ case, however, the relative ordering of the profiles is not straightforward, with the $x_a = 0.9$ and pure $x_a = 0$ cases being similar in magnitude at low fields. 

Figures~\ref{fig4}\,(e),\,(f),\,(h) and (i) demonstrate some combination of the effects described above, dependent on both the mixture structure and the relative electron scattering magnitudes of the components, leading to complex behaviour. In Figure~\ref{fig4}\,(e), the profiles intersect at approximately $E/n_0 = 0.5$ Td, i.e., a lower field than that observed in Figure~\ref{fig4}\,(b). In Figure~\ref{fig4}\,(h) the profiles demonstrate monotonic trends for the selection of density fractions displayed. In general, increasing $\alpha$, $\beta$ tends to decrease the drift velocity, while increasing $x_b$ has a non-monotonic effect.

\begin{table}[t]
    \centering
    \caption{Variations of transport properties (defined in equations~\eqref{eq:W}--\eqref{eq:DLT}) with percentage species $a$, for select reduced electric fields, $E/n_0$. The parameters of species $a$ and $b$ are set according to the model given in Figure~\ref{fig0}, with the values of $\alpha = \frac{\sigma_b}{\sigma_a}=4$, $\beta = \frac{d_b}{d_a}=2$ and $\Phi=0.4$.}
    \begin{tabular}{llllll}
    \hline
    \hline
    $E/n_0$ &  & $\epsilon$ & $W$ & $n_0D_T$ & $n_0D_L$ \\
    {[Td]} & $x_a\ \ \ \ $ & [eV]$\ \ \ \ \ $ & [$10^3$ms$^{-1}$] & [$10^{24}$m$^{-1}$s$^{-1}$] & [$10^{24}$m$^{-1}$s$^{-1}$] \\
    \hline    
    0.1 & 1 & 0.2966 & 17.91 & 40.66 & 19.01 \\
     & 0.75 & 0.1130 & 4.219 & 3.607 & 1.836 \\
     & 0.5 & 0.1276 & 5.058 & 4.909 & 2.426 \\ 
     & 0.25 & 0.1381 & 5.688 & 6.039 & 2.837 \\ 
     & 0 & 0.1451 & 6.119 & 6.930 & 3.051 \\ 
    1 & 1 & 2.476 & 42.74 & 90.31 & 27.37 \\
     & 0.75 & 1.254 & 15.61 & 14.46 & 7.536 \\
     & 0.5 & 1.190 & 14.25 & 13.99 & 4.559 \\ 
     & 0.25 & 1.118 & 12.91 & 13.02 & 3.204 \\ 
     & 0 & 1.055 & 11.80 & 12.01 & 2.479 \\
    10 & 1 & 10.44 & 36.08 & 67.24 & 3.163 \\
     & 0.75 & 5.670 & 14.59 & 13.34 & 1.322 \\
     & 0.5 & 4.564 & 10.515 & 9.301 & 0.7223 \\ 
     & 0.25 & 3.986 & 8.558 & 7.500 & 0.4498 \\ 
     & 0 & 3.628 & 7.415 & 6.394 & 0.3074 \\
    100 & 1 & 60.96 & 52.78 & 28.31 & 11.86 \\
     & 0.75 & 46.005 & 34.62 & 12.53 & 5.928 \\
     & 0.5 & 38.48 & 26.48 & 8.007 & 3.800 \\ 
     & 0.25 & 33.74 & 21.74 & 5.790 & 2.738 \\ 
     & 0 & 30.41 & 18.61 & 4.494 & 2.112 \\
     \hline
    \end{tabular}
    \label{tab1}
\end{table}

The values of the transport coefficients ($W$, $\epsilon$, $n_0D_L$, $n_0D_T$) for select values of the reduced electric field $E/n_0$ strength and density fraction $x_a$, for set $\alpha = 4$, $\beta = 2$ and $\Phi=0.4$ are enumerated in Table~\ref{tab1}, for benchmark purposes.  The pure case, $x_a = 1$, has been detailed previously~\cite{WhitRobs11,TattersallEtal2015,Boyl17}.Increasing the electric field strength tends to increase the corresponding transport properties, while decreasing $x_a$ tends to decrease the decrease the corresponding transport properties, as expected from the previous discussion.

The multi-term solution of Boltzmann's equation~\eqref{eq:Boltz} captures the anisotropy in the scattering process through the partial cross-sections~\eqref{eq:sig_l}--\eqref{eq:Sig_l}, and the number $l_{max}+1$ of terms retained in the expansions~\eqref{eq:F}--\eqref{eq:FT}. In practice, one simply increases the value of $l_{max}$ employed in the calculations until some desired level of convergence in the transport properties is reached. The convergence of the transport coefficients ($W$, $\epsilon$, $n_0D_L$, $n_0D_T$) with $l_{max}$ for select values of the reduced electric field $E/n_0$, for set $\alpha = 4$, $\beta = 2$, $\Phi=0.4$ and $x_a = 0.5$ is shown in Table~\ref{tab2}, for benchmark purposes. For the chosen fields, a two-term approximation $l_{max}=1$ is adequate to evaluate the mean energy and drift velocity to 4 significant figures. The two-term approximation for the transverse and longitudinal diffusion coefficients is also very reasonable, being accurate to within $2\%$  and $0.2\%$ of the converged values, respectively. Only $l_{max}=5$ is required for all transport properties to converge to 4 significant figures. The impact of anisotropic scattering via coherent elastic scattering, for the benchmark system, is captured well in low-order approximations. For anisotropic scattering, and when inelastic processes are present, this cannot be known \textit{a priori}~\cite{WhitEtal03,PetrEtal09}.

\begin{table}[t]
    \centering
    \caption{Convergence of transport properties (defined in equations~\eqref{eq:W}--\eqref{eq:DLT}) with $l_{max}$, for select reduced electric fields $E/n_0$. The parameters of species $a$ and $b$ are set according to the model given in Figure~\ref{fig0}, with the values of $\alpha = \frac{\sigma_b}{\sigma_a}=4$, $\beta = \frac{d_b}{d_a}=2$, $\Phi=0.4$ and $x_a=0.5$.}
    \begin{tabular}{llllll}
    \hline
    \hline
    $E/n_0$ &  & $\epsilon$ & $W$  & $n_0D_T$ & $n_0D_L$ \\
    {[Td]} & $l_{max}$ & [eV]$\ \ \ \ \ \ $ & [$10^3$ms$^{-1}$] & [$10^{24}$m$^{-1}$s$^{-1}$] & [$10^{24}$m$^{-1}$s$^{-1}$] \\
    \hline   
    0.1 & 1 & 0.1276 & 5.058 & 4.922 & 2.422 \\
     & 3 & 0.1276 & 5.058 & 4.909 & 2.426 \\
     & 5 & 0.1276 & 5.058 & 4.909 & 2.426 \\ 
    1 & 1 & 1.190 & 14.26 & 14.03 & 4.554 \\
     & 3 & 1.190 & 14.25 & 13.99 & 4.559 \\ 
     & 5 & 1.190 & 14.25 & 13.99 & 4.559 \\ 
    10 & 1 & 4.564 & 10.515 & 9.453 & 0.7218 \\
     & 3 & 4.564 & 10.515 & 9.303 & 0.7223 \\
     & 5 & 4.564 & 10.515 & 9.302 & 0.7223 \\
    100 & 1 & 38.48 & 26.48 & 8.143 & 3.800 \\
     & 3 & 38.48 & 26.48 & 8.012 & 3.800 \\
     & 5 & 38.48 & 26.48 & 8.007 & 3.800 \\ 
     \hline
    \end{tabular}
    \label{tab2}
\end{table} 

\subsection{Blanc's law approximation for mixtures}\label{sec:BlancsLaw}

We end this section with a discussion on Blanc's law for approximating mixtures from properties of the respective components, and its application to the binary liquid mixtures considered here. Blanc's law~\cite{Blanc1908}, which was developed for dilute gas mixtures, has the form
\begin{align}
    \frac{1}{K_{mix}} \approx x_a\frac{1}{K_a} + x_b\frac{1}{K_b}, \label{eq:BlancsLaw}
\end{align}
where $K = W/E$ is the electron mobility, such that $K_a$ and $K_b$ are the mobilities corresponding to an electric field $E$ for the pure species $a$ and $b$, respectively, and where $K_{mix}$ is the mixture mobility at that same field. This expression can be arrived at from the perspective of momentum-transfer-theory~\cite{RobsWhitHild17} which implies that this relation is exact when the energy-distribution functions of the pure and mixture systems are the same, and approximately correct when the mean energy of the pure and mixture systems are similar. Although Blanc's law was proposed in the early 1900s, it is still routinely used today, despite the well-documented limitations~\cite{SasicEtal2005,WangBrunt1997}. We briefly juxtapose the mixture results obtained using a Blanc's law against those using the full kinetic treatment, to highlight the necessity of the latter.

\begin{figure}[t]
\centering
\centering
    \includegraphics[width=1.0\linewidth, angle=0]{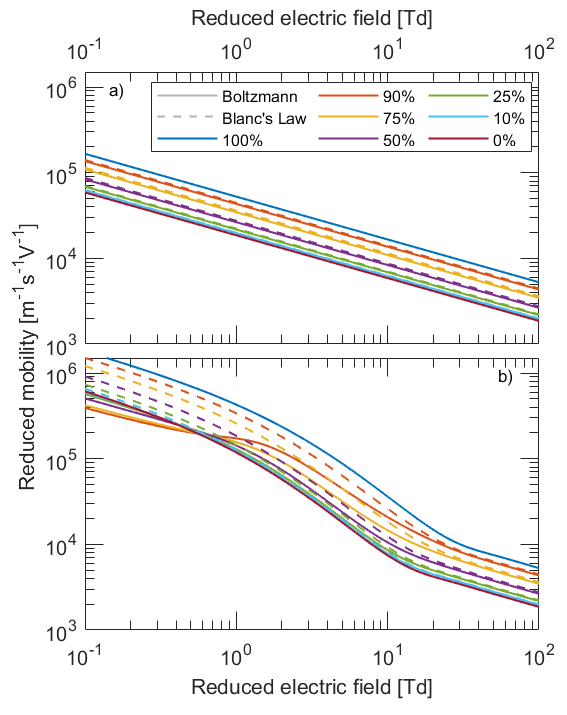}
    \caption{Comparison of reduced mobility, $n_0K$, for hard-sphere mixtures calculated with Boltzmann's equation~\eqref{eq:Boltz}, and those calculated using Blanc's Law ~\eqref{eq:BlancsLaw} from the pure components. The parameters of species $a$ and $b$ are set according to the model given in Figure~\ref{fig0}, with $\alpha = \frac{\sigma_b}{\sigma_a}=4$, $\beta = \frac{d_b}{d_a}=2$ and $\Phi=0.4$. \textbf{a)} Hard-sphere mixture in the binary-scattering limit. \textbf{b)} Hard-sphere coherent liquid mixture.}
    \label{fig5}
\end{figure} 

Figure~\ref{fig5} compares the reduced mobility for select mixture density fractions calculated via the solution of Boltzmann's equation~\ref{eq:Boltz}, with the application of Blanc's law using the pure $x_a = 1$ and $x_a=0$ mobilities (again, calculated using Boltzmann's equation). Figure~\ref{fig5}(a) shows the binary-scattering case, with electron-scattering cross-section described by hard-sphere constants, while Figure~\ref{fig5}(b) shows the liquid case, where the electron scattering cross-sections demonstrate complex structure-modified energy-dependencies. The profiles in Figure~\ref{fig5}(a) demonstrate simple power-law behaviour which appears linear in the log-log scale, and it can be seen that Blanc's law does a very good job of approximating the mobilities of the intermediate mixtures. In contrast, the profiles in Figure~\ref{fig5}(b) demonstrate the oscillatory, non-monotonic behaviour we expect from transport in structured media, and it is clear that Blanc's law does not satisfactorily approximate the mixture mobilities in this case. The agreement is at its worst at low fields, and at its best at high fields, which reflects the influence of structure. Garland and co-authors~\cite{GarlandEtal2018} investigated the generalization of Blanc's law to the interface between gas- and liquid-phase mediums, and it is expected a similar generalization for liquid mixtures is required here, which is left for future work.

\FloatBarrier

\section{Summary\label{sec:Summary}}

In this work, the formalism of Lekner and Cohen~\cite{Lekner1967,CohenLekner1967} to calculate the scattering cross-sections and transport properties of electrons in simple liquids has been extended to simple liquid mixtures. 

A kinetic equation collision operator has been developed to describe the low-energy coherent elastic scattering of electrons from the collective scattering centres of a binary fluid mixture, when the de Broglie wavelength of the electron is comparable to the interatomic spacing of the background medium. The extension to $m$-component mixtures should prove straightforward. This collision operator involves the angle-integrated partial static structure factors representing the self- and cross-correlations between the two species, $a$ and $b$, that make up the background medium. It was noted that the momentum-transfer scattering cross-section for the dense mixture must be treated as a single entity to preserve positivity, i.e., the separation into cross-sections `belonging' to species $a$ and $b$ individually can allow one of the components to violate positivity.

The analytic model of Percus and Yevick~\cite{PercusYevick1958} generalised to a multi-component mixture of hard-spheres~\cite{Lebowitz1964,Hiroike1969} was employed to characterise the structure effects of the background medium. The influence of structure properties including the relative size of components a and b, relative number density of species $a$ and $b$, and the total packing factor on the scattering cross-sections and associated transport properties ($W$, $\epsilon$, $n_0D_L$, $n_0D_T$) was investigated. For the benchmark system analysed, the second species featured a larger momentum-transfer cross-section for the electron scattering, and thus as the density fraction of the second species increased, the total momentum-transfer increased leading to an overall decrease in the corresponding transport properties. This was most prevalent at electric fields that corresponded to the low-energy region of the cross-sections. Similiarly, increasing the packing factor $\Phi$ tended to decrease the momentum-transfer at low energies, leading to an enhancement of the transport properties at low field, as is the case for electron transport in pure liquids. By varying the relative electron-scattering magnitudes and relative hard-sphere diameters in the liquid mixture independently, the influence of each on the transport was isolated. The drift velocity as a function of electric field demonstrated complicated non-monotonic behaviour as the density fractions were varied.

The complex energy-dependence of the structure-modified cross-sections also ensures that the straightforward application of Blanc's law~\cite{Blanc1908} for calculating the reduced mobility of mixtures using the reduced mobility of the pure liquids is insufficient, despite being a satisfactory approximation in the binary-scattering regime. This suggests that a more sophisticated approach, such as that employed in Ref. \cite{GarlandEtal2018}, is required.

The formalism here describing the coherent scattering and transport effects in mixtures must be supplemented by a mixture model of the modifications to the individual scattering potentials~\cite{Lekner1967}. Future work will be to consider real mixtures systems, such as noble liquid systems~\cite{Borghesani1993,Borghesani1997}, particularly those relevant to liquid dark matter detector development~\cite{EFCA2021}.

\begin{acknowledgments}
The authors wish to acknowledge helpful discussions with D. L. Muccignat. The authors would also like to thank the Australian Research Council through its Discovery program (DP190100696,DP220101480).
\end{acknowledgments}

\appendix

\section{Derivation of differential scattering
cross-sections in liquid mixture systems\label{sec:DCS}}

In this section we will follow and adapt the derivation given by Robson, White and Hildebrandt~\cite{RobsWhitHild17}. For simplicity, we will only consider non-polar molecules/atoms (for the extension to including the influence of a permanent dipole moment, see Ref.~\cite{RobsWhitHild17}).

Consider an incoming plane wave, $\psi_P$, of amplitude $A_0$,
\begin{align}
    \psi_P = A_0e^{i(\mathbf{k}\cdot \mathbf{r} - \omega t)}
\end{align}
incident on a volume $V$, scattered into an outgoing spherical wave centred on $\mathbf{r}$ by molecules of the medium located within a volume $d\mathbf{r}$ of position $\mathbf{r}$. The angle between the wave vectors $\mathbf{k}$ and $\mathbf{k'}$ of the incoming and outgoing waves respectively is $\theta$, where we employ the usual correspondence between wave and particle properties, i.e.,
\begin{align}
    \mathbf{p} &= \hbar\mathbf{k}, \\
    \epsilon &= \hbar\omega = \frac{\hbar^2 k^2}{2m_e}.
\end{align}
The amplitude of the outgoing spherical wave, $A'$, will have a contribution due to each component, a and b. of the binary mixture, with each component being proportional to; i) a factor ${f}_{a.b}(\mathbf{k},\mathbf{k'})$ accounting for scattering from single centres, ii) the local fluctuations $\eta_{a,b}(\mathbf{r},t)$ in the density of scatterers, iii) the amplitude of the incoming wave, and iv) the differential volume $d\mathbf{r}$ from which the scattering takes place. The contribution to the scattered wave, measured at some distance $R \gg r$, due to scattering from molecules within $d\mathbf{r}$ of $\mathbf{r}$ is  
\begin{align}
     d\psi_S &= C\left[{f}_{a}(\mathbf{k},\mathbf{k'})\eta_{a}(\mathbf{r},t) + {f}_{b}(\mathbf{k},\mathbf{k'})\eta_{b}(\mathbf{r},t)\right] \nonumber \\
    &\ \ \ \ \ \ \ \times A_0e^{i(\mathbf{k}\cdot \mathbf{r} - \omega t)}\frac{e^{ik'|\mathbf{R}-\mathbf{r}|}}{|\mathbf{R}-\mathbf{r}|}d\mathbf{r},
\end{align}
where $C$ is a constant of proportionality. The contribution to the scattered amplitude from all molecules in the volume $V$ is 
\begin{align}
    \frac{\psi_S}{CA_0} &= {f}_{a}(\mathbf{k},\mathbf{k'})\!\!\int_V \!\! \eta_{a}(\mathbf{r},t) e^{i(\mathbf{k}\cdot \mathbf{r} - \omega t)}\frac{e^{ik'|\mathbf{R}-\mathbf{r}|}}{|\mathbf{R}-\mathbf{r}|} d\mathbf{r} \nonumber\\
    &\ \ + {f}_{b}(\mathbf{k},\mathbf{k'})\!\!\int_V \!\! \eta_{b}(\mathbf{r},t) e^{i(\mathbf{k}\cdot \mathbf{r} - \omega t)}\frac{e^{ik'|\mathbf{R}-\mathbf{r}|}}{|\mathbf{R}-\mathbf{r}|}d\mathbf{r}.
\end{align}
Since $R \gg r$, then $k'|\mathbf{R}-\mathbf{r}| \approx kR - \mathbf{k'}\cdot\mathbf{r}$, and hence
\begin{align}
    \psi_S &\approx A'\frac{e^{ik'R}}{R} =  \left(A'_{a} + A'_{b}\right)\frac{e^{ik'R}}{R},
\end{align}
where
\begin{align}
    \frac{A'_{a,b}}{CA_0} &= {f}_{a,b}(\mathbf{k},\mathbf{k'})\!\!\int_V \!\! \eta_{a,b}(\mathbf{r},t) e^{i\left[(\mathbf{k}-\mathbf{k'})\cdot \mathbf{r} - \omega t\right]}\,d\mathbf{r}.
\end{align}
Since the properties of the scattering centres in the medium mixture are subject to random fluctuations, then these amplitudes themselves are also randomly fluctuating quantities. A measure of the averaged properties of the fluctuations is provided by the correlation function~\cite{HansenMcDonald1973}, $B(\tau)$,
\begin{align}
    B(\tau) &= \langle A'(t)A'(t-\tau)^*\rangle, \\
    &= \langle A_{a}'(t)A_{a}'(t-\tau)^*\rangle  + \langle A_{a}'(t)A_{b}'(t-\tau)^*\rangle   \nonumber \\
    &\ \ \ + \langle A_{b}'(t)A_{a}'(t-\tau)^*\rangle   + \langle A_{b}'(t)A_{b}'(t-\tau)^*\rangle, 
\end{align}
where, 
\begin{align}
    &\frac{\langle A_{n}'(t)A_{m}'(t-\tau)^*\rangle}{|C|^2|A_0^2|}
    = {f}_{a}(\mathbf{k},\mathbf{k'}){f}_{a}^*(\mathbf{k},\mathbf{k'}) \!\! \nonumber\\
    & \ \ \ \times \!\!\! \int \!\! d\mathbf{r}_1 \!\! \int \!\! d\mathbf{r}_2 \, \langle  \eta_{a}(\mathbf{r}_1,t)\eta_{a}^*(\mathbf{r}_2,t\!-\!\tau) \rangle e^{i[ \Delta\mathbf{k}\cdot\Delta\mathbf{r}-\omega \tau]}, \label{eq:corr}
\end{align} 
and $\Delta \mathbf{k} = \mathbf{k}-\mathbf{k'}$ and $\Delta \mathbf{r} = \mathbf{r}_1-\mathbf{r}_2$.
The spectral power density is related to the correlation functions via the relation
\begin{align}
    \Phi(\omega') = \frac{1}{2\pi} \int_{-\infty}^{\infty}\!d\tau\, e^{i\omega'\tau} B(\tau),
\end{align}
and inverse relation
\begin{align}
    B(\tau) = \int_{-\infty}^{\infty}\!d\tau\, e^{-i\omega'\tau} \Phi(\tau).
\end{align}
The spectral power density can thus be written as,
\begin{align}
    &\frac{\Phi(\omega')}{(2\pi)^3V|C|^2|A_0^2|} \!=
    {f}_{a}(\mathbf{k},\mathbf{k'}) {f}_{a}^*(\mathbf{k},\mathbf{k'}) {S}_{aa}(\Delta \mathbf{k}, \Delta \omega) \nonumber \\
    &\ \ \ \ \ \ \ \ \ \ \ \ \ \ \ \ \ \ \ \  + {f}_{a}(\mathbf{k},\mathbf{k'}) {f}_{b}^*(\mathbf{k},\mathbf{k'}) {S}_{ab}(\Delta \mathbf{k}, \Delta \omega) \nonumber \\
    &\ \ \ \ \ \ \ \ \ \ \ \ \ \ \ \ \ \ \ \  + {f}_{b}(\mathbf{k},\mathbf{k'}) {f}_{a}^*(\mathbf{k},\mathbf{k'}) {S}_{ba}(\Delta \mathbf{k}, \Delta \omega) \nonumber\\
    &\ \ \ \ \ \ \ \ \ \ \ \ \ \ \ \ \ \ \ \ + {f}_{b}(\mathbf{k},\mathbf{k'}) {f}_{b}^*(\mathbf{k},\mathbf{k'}) {S}_{bb}(\Delta \mathbf{k}, \Delta\omega),
\end{align}
where $\Delta\omega = \omega - \omega'$, and the ${S}_{mn}(K, \Omega)$ are dynamic structure factors~\cite{HansenMcDonald1973},
\begin{align}
    {S}_{mn}(\Delta \mathbf{k}, \Delta \omega) \! &= \! \frac{1}{(2\pi)^4V} \! \int \!\! d\mathbf{r}_1 \!\! \int \!\! d\mathbf{r}_2 \!\! \int_{-\infty}^{\infty} \!\! \! \! d\tau\, G_{m,n}(\mathbf{r}_1,\! \mathbf{r}_2,\!\tau),
\end{align}
with $G_{mn}$ representing the full time-space correlations,
\begin{align}
    G_{mn}(\mathbf{r}_1,\! \mathbf{r}_2,\!\tau) \! &= \! \langle  \eta_{m}(\mathbf{r}_1,t)\eta_{n}^*(\mathbf{r}_2,t\!-\!\tau) \rangle e^{i[ \Delta \mathbf{k}\cdot\Delta\mathbf{r}-\Delta \omega \tau]}.
\end{align} 

The bulk differential cross-section for all molecules in the region, per unit volume, is defined via,
\begin{align}
    \frac{d\sigma}{d\Omega_{k'}} &= \frac{\langle |A'(t)|^2\rangle}{V|A_0|^2} \\
    &= \frac{B(0)}{V|A_0|^2} = \frac{1}{V|A_0|^2} \int_{-\infty}^{\infty} d\omega' \Phi(\omega')  \\
    &= {f}_{a}(\mathbf{k},\mathbf{k'}) {f}_{a}^*(\mathbf{k},\mathbf{k'})  \int_{-\infty}^{\infty} d\omega' {S}_{aa}(\Delta \mathbf{k}, \Delta \omega) \nonumber \\
    &+ {f}_{a}(\mathbf{k},\mathbf{k'}) {f}_{b}^*(\mathbf{k},\mathbf{k'})  \int_{-\infty}^{\infty} d\omega' {S}_{ab}(\Delta \mathbf{k}, \Delta \omega) \nonumber \\
    &+ {f}_{b}(\mathbf{k},\mathbf{k'}) {f}_{a}^*(\mathbf{k},\mathbf{k'})  \int_{-\infty}^{\infty} d\omega' {S}_{ba}(\Delta \mathbf{k}, \Delta \omega) \nonumber\\
    &+ {f}_{b}(\mathbf{k},\mathbf{k'}) {f}_{b}^*(\mathbf{k},\mathbf{k'})  \int_{-\infty}^{\infty} d\omega' {S}_{bb}(\Delta \mathbf{k}, \Delta\omega), \label{eq:DCS_deriv1}
\end{align}
where the  normalisation constant has been chosen such that $(2\pi)^3|C|^2=1$. Noting that the static structure factor is related to the dynamic structure factor via,
\begin{align}
    S_{mn}(|\Delta \mathbf{k}|) = \int_{-\infty}^{\infty} d\omega' {S}_{mn}(\Delta \mathbf{k}, \Delta \omega),
\end{align}
and identifying
\begin{align}
     f_a(\mathbf{k},\mathbf{k'})f_b^*(\mathbf{k},\mathbf{k'}) = n_0\sqrt{ \left(\frac{d\sigma}{d\Omega_{k'}} \right)^{lab}_a\!\! \left(\frac{d\sigma}{d\Omega_{k'}} \right)^{lab}_b}, \label{eq:labframe}
\end{align}
for binary scattering off an individual molecule in the laboratory frame, then equation~\eqref{eq:DCS_deriv1} can be written as
\begin{align}
     \frac{d\sigma}{d\Omega_{k'}} &= n_0\!\left(\frac{d\sigma}{d\Omega_{k'}} \right)^{lab}_a\!\!\!{S}_{aa}(|\Delta \mathbf{k}|) \nonumber \\
     &\ \ \ + n_0\! \left(\frac{d\sigma}{d\Omega_{k'}} \right)^{lab}_b\!\!\!{S}_{bb}(|\Delta \mathbf{k}|) \nonumber \\
     &\ \ \ + 2n_0\sqrt{ \left(\frac{d\sigma}{d\Omega_{k'}} \right)^{lab}_a\!\! \left(\frac{d\sigma}{d\Omega_{k'}} \right)^{lab}_b}{S}_{ab}(|\Delta \mathbf{k}|), \label{eq:DCSapp}
\end{align}
where we have used the fact that $S_{ab} = S_{ba}$, and where the total number density $n_0$ of scattering centres has been made explicit for consistency with the normalisation of the structure factors to unit particle number. Equation~\eqref{eq:DCSapp} then represents the total differential scattering cross-section with respect to binary mixtures including structure effects.

Noting the definition and properties of the partial static structure factors discussed in section~\ref{sec:Struc}, in the limit of large values of $|\Delta \mathbf{k}|$, $S_{aa} \rightarrow x_a, S_{bb} \rightarrow x_b$ and $S_{ab} \rightarrow 0$, such that 
\begin{align}
     \lim_{|\Delta \mathbf{k}|\to\infty}  \frac{d\sigma}{d\Omega_{k'}} \! &=\! x_a n_0\! \left(\frac{d\sigma}{d\Omega_{k'}} \right)^{lab}_a \! \! \! + x_b n_0\! \left(\frac{d\sigma}{d\Omega_{k'}} \right)^{lab}_b\!\!\!,
\end{align}
i.e. the binary-scattering result is recovered, as required. Similarly, in the case of $b \rightarrow a$, $\sigma_b \rightarrow \sigma_a$ and $S_{aa} + 2S_{ab} + S_{bb} \rightarrow S_a$, where $S_a$ represents the structure factor for a pure liquid of species $a$, recovering the pure liquid case, as required.

\section{Note on the collision operator \label{sec:ColOpApp}}

In the section we describe the considerations that need to be made in deriving the elastic collision operator for mixtures, accounting for coherent scattering effects.

When considering the transport of light particles, such as electrons, one can assume that the background medium remains essentially in undisturbed  thermal equilibrium, such that the energy distribution of the background particles is a stationary Maxwellian, and the collision operator $J(f)$ in equation~\eqref{eq:Boltz} becomes a linear operator~\cite{RobsWhitHild17}. The collision operator corresponding to the three terms in the differential scattering cross-section equation~\eqref{eq:DCSapp} can thus be treated separately. 

Following Ref.~\cite{RobsWhitHild17}, in order to derive the elastic collision operators, the lab-frame differential scattering cross-sections and dynamic structure factors in equations~\eqref{eq:DCS_deriv1}--\eqref{eq:DCSapp} must be expanded in terms of ${\Delta \omega = (k^2-k'^2)/2m_e\approx k(k-k')/m_e}$, so as to evaluate the collision integrals. The individual lab-frame differential scattering cross-sections, and each of the partial dynamic structure factors already have the same form as those treated in Ref.~\cite{RobsWhitHild17}, and thus it is only the square root term that remains to be evaluated. Starting from the following expansion in terms of $\Delta \omega$,
\begin{align}
    \! \left(\frac{d\sigma}{d\Omega_{k'}} \right)^{lab} \! \! \! \! =  \sigma(\omega,\theta) \! - \! \frac{\Delta\omega}{2}\frac{\partial}{\partial \omega}\sigma(\omega,\theta) + \mathcal{O}\!\left([\Delta\omega]^2\right),
\end{align}
it follows that 
\begin{align}
    &\sqrt{ \left(\frac{d\sigma}{d\Omega_{k'}} \right)^{lab}_a\! \left(\frac{d\sigma}{d\Omega_{k'}} \right)^{lab}_b} \nonumber \\
    &\ = \left(\!1 \! - \! \frac{\Delta\omega}{2}\frac{\partial}{\partial \omega}\!\right)\sqrt{\sigma_a(\omega,\theta)\sigma_b(\omega,\theta)} + \mathcal{O}\!\left([\Delta\omega]^2\right), \\
    &\ \approx \sigma_{ab}(\omega,\theta) - \frac{\Delta\omega}{2}\frac{\partial}{\partial \omega}\sigma_{ab}(\omega,\theta),
\end{align}
where defining $\sigma_{ab} \equiv \sqrt{\sigma_a(\omega,\theta)\sigma_b(\omega,\theta)}$ makes it clear that each of the three terms in equation~\eqref{eq:DCSapp} have the exact same form, to first order in $\Delta\omega$, as the pure liquid case~\cite{RobsWhitHild17}, and thus the collision operators also have the same form. Thus, for elastic collisions involving light particles, the $J_0$ operator reduces to the usual binary-scattering case (see Ref.~\cite{RobsWhitHild17} for details), which for a binary mixture is
    \begin{align}
    J_0 (f_0) &= \frac{m_e}{m_a}\frac{1}{v^2}\left[v\nu^a_1(v)\left(vf_0 + \frac{k_bT}{m_e}\frac{d}{dv}f_0\right)\right] \nonumber \\
    &+ \frac{m_e}{m_b}\frac{1}{v^2}\left[v\nu^b_1(v)\left(vf_0 + \frac{k_bT}{m_e}\frac{d}{dv}f_0\right)\right],
\end{align} 
where the superscripts $a$ and $b$ indicate the two different medium species, and where
\begin{align}
    \nu_1(v) &= n_0 v\,2\pi\!\!\int_0^{2\pi} \!\!\! \sigma(v,\theta)[1\!-\!\cos\theta]\sin\theta\, d\theta,
\end{align}
is the binary-scattering momentum-transfer collision frequency. The structure of the fluid plays a direct role in the vector and higher rank tensor collision operator components, $J_l$, which, for a binary mixture, are then
\begin{align}
     J_l (f_l) &= \tilde{\nu}_l(v) f_l, 
 \end{align}
for $l \geq 1$, where 
\begin{align}
    \tilde{\nu}_l(v) &= n_0v2\pi\int_0^{2\pi}\Sigma(v,\theta)[1-P_l(\cos\theta)]\sin\theta d\theta
\end{align}
and where $\Sigma(v,\theta)$ is structure-modified cross-section, 
\begin{align}
    \Sigma(v,\theta)
    &= n_0\sigma_a(v,\theta)S_{aa}\left(\frac{2m_ev}{\hbar}\sin{\frac{\theta}{2}}\right)   \nonumber \\ 
    &+ n_0\sigma_b(v,\theta) S_{bb}\left(\frac{2m_ev}{\hbar}\sin{\frac{\theta}{2}}\right) \nonumber \\
    &+ 2n_0 \sqrt{\sigma_a(v,\theta)\sigma_b(v,\theta)} S_{ba}\!\left(\frac{2m_ev}{\hbar}\sin{\frac{\theta}{2}}\right).
\end{align}
In the limit of large argument values, $S_{aa} \rightarrow x_a, S_{bb} \rightarrow x_b$ and $S_{ab} \rightarrow 0$, such that the usual dilute gas mixture case is recovered, as required. In the case of $b \rightarrow a$, $\sigma_b \rightarrow \sigma_a$ and $S_{aa} + 2S_{ab} + S_{bb} \rightarrow S_a$, where $S_a$ represents the structure factor for a pure liquid of species $a$, as required.

\bibliography{apssamp}

\providecommand{\noopsort}[1]{}\providecommand{\singleletter}[1]{#1}%
\begin{thebibliography}{31}%
\makeatletter
\providecommand \@ifxundefined [1]{%
 \@ifx{#1\undefined}
}%
\providecommand \@ifnum [1]{%
 \ifnum #1\expandafter \@firstoftwo
 \else \expandafter \@secondoftwo
 \fi
}%
\providecommand \@ifx [1]{%
 \ifx #1\expandafter \@firstoftwo
 \else \expandafter \@secondoftwo
 \fi
}%
\providecommand \natexlab [1]{#1}%
\providecommand \enquote  [1]{``#1''}%
\providecommand \bibnamefont  [1]{#1}%
\providecommand \bibfnamefont [1]{#1}%
\providecommand \citenamefont [1]{#1}%
\providecommand \href@noop [0]{\@secondoftwo}%
\providecommand \href [0]{\begingroup \@sanitize@url \@href}%
\providecommand \@href[1]{\@@startlink{#1}\@@href}%
\providecommand \@@href[1]{\endgroup#1\@@endlink}%
\providecommand \@sanitize@url [0]{\catcode `\\12\catcode `\$12\catcode `\&12\catcode `\#12\catcode `\^12\catcode `\_12\catcode `\%12\relax}%
\providecommand \@@startlink[1]{}%
\providecommand \@@endlink[0]{}%
\providecommand \url  [0]{\begingroup\@sanitize@url \@url }%
\providecommand \@url [1]{\endgroup\@href {#1}{\urlprefix }}%
\providecommand \urlprefix  [0]{URL }%
\providecommand \Eprint [0]{\href }%
\providecommand \doibase [0]{https://doi.org/}%
\providecommand \selectlanguage [0]{\@gobble}%
\providecommand \bibinfo  [0]{\@secondoftwo}%
\providecommand \bibfield  [0]{\@secondoftwo}%
\providecommand \translation [1]{[#1]}%
\providecommand \BibitemOpen [0]{}%
\providecommand \bibitemStop [0]{}%
\providecommand \bibitemNoStop [0]{.\EOS\space}%
\providecommand \EOS [0]{\spacefactor3000\relax}%
\providecommand \BibitemShut  [1]{\csname bibitem#1\endcsname}%
\let\auto@bib@innerbib\@empty
\bibitem [{\citenamefont {Cohen}\ and\ \citenamefont {Lekner}(1967)}]{CohenLekner1967}%
  \BibitemOpen
  \bibfield  {author} {\bibinfo {author} {\bibfnamefont {M.~H.}\ \bibnamefont {Cohen}}\ and\ \bibinfo {author} {\bibfnamefont {J.}~\bibnamefont {Lekner}},\ }\bibfield  {title} {\bibinfo {title} {Theory of hot electrons in gases, liquids, and solids},\ }\href {https://doi.org/10.1103/PhysRev.158.305} {\bibfield  {journal} {\bibinfo  {journal} {Phys. Rev.}\ }\textbf {\bibinfo {volume} {158}},\ \bibinfo {pages} {305} (\bibinfo {year} {1967})}\BibitemShut {NoStop}%
\bibitem [{\citenamefont {Lebowitz}(1964)}]{Lebowitz1964}%
  \BibitemOpen
  \bibfield  {author} {\bibinfo {author} {\bibfnamefont {J.~L.}\ \bibnamefont {Lebowitz}},\ }\bibfield  {title} {\bibinfo {title} {Exact solution of generalized {P}ercus-{Y}evick equation for a mixture of hard spheres},\ }\href {https://doi.org/10.1103/PhysRev.133.A895} {\bibfield  {journal} {\bibinfo  {journal} {Phys. Rev.}\ }\textbf {\bibinfo {volume} {133}},\ \bibinfo {pages} {A895} (\bibinfo {year} {1964})}\BibitemShut {NoStop}%
\bibitem [{\citenamefont {Hiroike}(1969)}]{Hiroike1969}%
  \BibitemOpen
  \bibfield  {author} {\bibinfo {author} {\bibfnamefont {K.}~\bibnamefont {Hiroike}},\ }\bibfield  {title} {\bibinfo {title} {Ornstein-zernike relation for a fluid mixture with direct correlation functions of finite range},\ }\href {https://doi.org/10.1143/JPSJ.27.1415} {\bibfield  {journal} {\bibinfo  {journal} {Journal of the Physical Society of Japan}\ }\textbf {\bibinfo {volume} {27}},\ \bibinfo {pages} {1415} (\bibinfo {year} {1969})}\BibitemShut {NoStop}%
\bibitem [{\citenamefont {Colaleo}\ \emph {et~al.}(2021)\citenamefont {Colaleo}, \citenamefont {Ropelewski}, \citenamefont {Dehmelt}, \citenamefont {Liberti}, \citenamefont {Titov}, \citenamefont {Veloso}, \citenamefont {Monroe}, \citenamefont {Colijn}, \citenamefont {Ereditato}, \citenamefont {Botella}, \citenamefont {Lindner}, \citenamefont {Cartiglia}, \citenamefont {Pellegrini}, \citenamefont {Bortoletto}, \citenamefont {Contardo}, \citenamefont {Gregor}, \citenamefont {Kramberger}, \citenamefont {Pernegger}, \citenamefont {Harnew}, \citenamefont {Krizan}, \citenamefont {Adachi}, \citenamefont {Nappi}, \citenamefont {Joram}, \citenamefont {Schultz-Coulon}, \citenamefont {Demarteau}, \citenamefont {Doser}, \citenamefont {Braggio}, \citenamefont {Geraci}, \citenamefont {Graham}, \citenamefont {Grasselino}, \citenamefont {Russell}, \citenamefont {Withington}, \citenamefont {Ferrari}, \citenamefont {Poschl}, \citenamefont {Aleksa}, \citenamefont {Barney}, \citenamefont {Simon}, \citenamefont {Fatis},
  \citenamefont {Newbold}, \citenamefont {Vasey}, \citenamefont {Neufeld}, \citenamefont {Re}, \citenamefont {Taille}, \citenamefont {Weber}, \citenamefont {Hartmann}, \citenamefont {Riegler}, \citenamefont {Gargiulo}, \citenamefont {Resnati}, \citenamefont {Kate}, \citenamefont {Verlaat}, \citenamefont {Vos}, \citenamefont {Collot}, \citenamefont {Garutti}, \citenamefont {Brenner}, \citenamefont {Bakel}, \citenamefont {Gwenlan}, \citenamefont {Wiener}, \citenamefont {Appleby}, \citenamefont {Allport}, \citenamefont {Torre}, \citenamefont {Krammer}, \citenamefont {Sefkow}, \citenamefont {Shipsey}, \citenamefont {Jakobs}, \citenamefont {D'hondt}, \citenamefont {Rivkin}, \citenamefont {Colaleo}, \citenamefont {Ropelewski}, \citenamefont {Dehmelt}, \citenamefont {Liberti}, \citenamefont {Titov}, \citenamefont {Veloso}, \citenamefont {Monroe}, \citenamefont {Colijn}, \citenamefont {Ereditato}, \citenamefont {Botella}, \citenamefont {Lindner}, \citenamefont {Cartiglia}, \citenamefont {Pellegrini}, \citenamefont
  {Bortoletto}, \citenamefont {Contardo}, \citenamefont {Gregor}, \citenamefont {Kramberger}, \citenamefont {Pernegger}, \citenamefont {Harnew}, \citenamefont {Krizan}, \citenamefont {Adachi}, \citenamefont {Nappi}, \citenamefont {Joram}, \citenamefont {Schultz-Coulon}, \citenamefont {Demarteau}, \citenamefont {Doser}, \citenamefont {Braggio}, \citenamefont {Geraci}, \citenamefont {Graham}, \citenamefont {Grasselino}, \citenamefont {Russell}, \citenamefont {Withington}, \citenamefont {Ferrari}, \citenamefont {Poschl}, \citenamefont {Aleksa}, \citenamefont {Barney}, \citenamefont {Simon}, \citenamefont {Fatis}, \citenamefont {Newbold}, \citenamefont {Vasey}, \citenamefont {Neufeld}, \citenamefont {Re}, \citenamefont {Taille}, \citenamefont {Weber}, \citenamefont {Hartmann}, \citenamefont {Riegler}, \citenamefont {Gargiulo}, \citenamefont {Resnati}, \citenamefont {Kate}, \citenamefont {Verlaat}, \citenamefont {Vos}, \citenamefont {Collot}, \citenamefont {Garutti}, \citenamefont {Brenner}, \citenamefont {Bakel},
  \citenamefont {Gwenlan}, \citenamefont {Wiener}, \citenamefont {Appleby}, \citenamefont {Allport}, \citenamefont {Torre}, \citenamefont {Krammer}, \citenamefont {Sefkow}, \citenamefont {Shipsey}, \citenamefont {Jakobs}, \citenamefont {D'hondt},\ and\ \citenamefont {Rivkin}}]{EFCA2021}%
  \BibitemOpen
  \bibfield  {author} {\bibinfo {author} {\bibfnamefont {A.}~\bibnamefont {Colaleo}}, \bibinfo {author} {\bibfnamefont {L.}~\bibnamefont {Ropelewski}}, \bibinfo {author} {\bibfnamefont {K.}~\bibnamefont {Dehmelt}}, \bibinfo {author} {\bibfnamefont {B.}~\bibnamefont {Liberti}}, \bibinfo {author} {\bibfnamefont {M.}~\bibnamefont {Titov}}, \bibinfo {author} {\bibfnamefont {J.}~\bibnamefont {Veloso}}, \bibinfo {author} {\bibfnamefont {J.}~\bibnamefont {Monroe}}, \bibinfo {author} {\bibfnamefont {A.-P.}\ \bibnamefont {Colijn}}, \bibinfo {author} {\bibfnamefont {A.}~\bibnamefont {Ereditato}}, \bibinfo {author} {\bibfnamefont {I.~G.}\ \bibnamefont {Botella}}, \bibinfo {author} {\bibfnamefont {M.}~\bibnamefont {Lindner}}, \bibinfo {author} {\bibfnamefont {N.}~\bibnamefont {Cartiglia}}, \bibinfo {author} {\bibfnamefont {G.}~\bibnamefont {Pellegrini}}, \bibinfo {author} {\bibfnamefont {D.}~\bibnamefont {Bortoletto}}, \bibinfo {author} {\bibfnamefont {D.}~\bibnamefont {Contardo}}, \bibinfo {author} {\bibfnamefont
  {I.-M.}\ \bibnamefont {Gregor}}, \bibinfo {author} {\bibfnamefont {G.}~\bibnamefont {Kramberger}}, \bibinfo {author} {\bibfnamefont {H.}~\bibnamefont {Pernegger}}, \bibinfo {author} {\bibfnamefont {N.}~\bibnamefont {Harnew}}, \bibinfo {author} {\bibfnamefont {P.}~\bibnamefont {Krizan}}, \bibinfo {author} {\bibfnamefont {I.}~\bibnamefont {Adachi}}, \bibinfo {author} {\bibfnamefont {E.}~\bibnamefont {Nappi}}, \bibinfo {author} {\bibfnamefont {C.}~\bibnamefont {Joram}}, \bibinfo {author} {\bibfnamefont {H.-C.}\ \bibnamefont {Schultz-Coulon}}, \bibinfo {author} {\bibfnamefont {M.}~\bibnamefont {Demarteau}}, \bibinfo {author} {\bibfnamefont {M.}~\bibnamefont {Doser}}, \bibinfo {author} {\bibfnamefont {C.}~\bibnamefont {Braggio}}, \bibinfo {author} {\bibfnamefont {A.}~\bibnamefont {Geraci}}, \bibinfo {author} {\bibfnamefont {P.}~\bibnamefont {Graham}}, \bibinfo {author} {\bibfnamefont {A.}~\bibnamefont {Grasselino}}, \bibinfo {author} {\bibfnamefont {J.~M.}\ \bibnamefont {Russell}}, \bibinfo {author}
  {\bibfnamefont {S.}~\bibnamefont {Withington}}, \bibinfo {author} {\bibfnamefont {R.}~\bibnamefont {Ferrari}}, \bibinfo {author} {\bibfnamefont {R.}~\bibnamefont {Poschl}}, \bibinfo {author} {\bibfnamefont {M.}~\bibnamefont {Aleksa}}, \bibinfo {author} {\bibfnamefont {D.}~\bibnamefont {Barney}}, \bibinfo {author} {\bibfnamefont {F.}~\bibnamefont {Simon}}, \bibinfo {author} {\bibfnamefont {T.~T.~D.}\ \bibnamefont {Fatis}}, \bibinfo {author} {\bibfnamefont {D.}~\bibnamefont {Newbold}}, \bibinfo {author} {\bibfnamefont {F.}~\bibnamefont {Vasey}}, \bibinfo {author} {\bibfnamefont {N.}~\bibnamefont {Neufeld}}, \bibinfo {author} {\bibfnamefont {V.}~\bibnamefont {Re}}, \bibinfo {author} {\bibfnamefont {C.~d.~L.}\ \bibnamefont {Taille}}, \bibinfo {author} {\bibfnamefont {M.}~\bibnamefont {Weber}}, \bibinfo {author} {\bibfnamefont {F.}~\bibnamefont {Hartmann}}, \bibinfo {author} {\bibfnamefont {W.}~\bibnamefont {Riegler}}, \bibinfo {author} {\bibfnamefont {C.}~\bibnamefont {Gargiulo}}, \bibinfo {author}
  {\bibfnamefont {F.}~\bibnamefont {Resnati}}, \bibinfo {author} {\bibfnamefont {H.~T.}\ \bibnamefont {Kate}}, \bibinfo {author} {\bibfnamefont {B.}~\bibnamefont {Verlaat}}, \bibinfo {author} {\bibfnamefont {M.}~\bibnamefont {Vos}}, \bibinfo {author} {\bibfnamefont {J.}~\bibnamefont {Collot}}, \bibinfo {author} {\bibfnamefont {E.}~\bibnamefont {Garutti}}, \bibinfo {author} {\bibfnamefont {R.}~\bibnamefont {Brenner}}, \bibinfo {author} {\bibfnamefont {N.~V.}\ \bibnamefont {Bakel}}, \bibinfo {author} {\bibfnamefont {C.}~\bibnamefont {Gwenlan}}, \bibinfo {author} {\bibfnamefont {J.}~\bibnamefont {Wiener}}, \bibinfo {author} {\bibfnamefont {R.}~\bibnamefont {Appleby}}, \bibinfo {author} {\bibfnamefont {P.}~\bibnamefont {Allport}}, \bibinfo {author} {\bibfnamefont {S.~D.}\ \bibnamefont {Torre}}, \bibinfo {author} {\bibfnamefont {M.}~\bibnamefont {Krammer}}, \bibinfo {author} {\bibfnamefont {F.}~\bibnamefont {Sefkow}}, \bibinfo {author} {\bibfnamefont {I.}~\bibnamefont {Shipsey}}, \bibinfo {author} {\bibfnamefont
  {K.}~\bibnamefont {Jakobs}}, \bibinfo {author} {\bibfnamefont {J.}~\bibnamefont {D'hondt}}, \bibinfo {author} {\bibfnamefont {L.}~\bibnamefont {Rivkin}}, \bibinfo {author} {\bibfnamefont {A.}~\bibnamefont {Colaleo}}, \bibinfo {author} {\bibfnamefont {L.}~\bibnamefont {Ropelewski}}, \bibinfo {author} {\bibfnamefont {K.}~\bibnamefont {Dehmelt}}, \bibinfo {author} {\bibfnamefont {B.}~\bibnamefont {Liberti}}, \bibinfo {author} {\bibfnamefont {M.}~\bibnamefont {Titov}}, \bibinfo {author} {\bibfnamefont {J.}~\bibnamefont {Veloso}}, \bibinfo {author} {\bibfnamefont {J.}~\bibnamefont {Monroe}}, \bibinfo {author} {\bibfnamefont {A.-P.}\ \bibnamefont {Colijn}}, \bibinfo {author} {\bibfnamefont {A.}~\bibnamefont {Ereditato}}, \bibinfo {author} {\bibfnamefont {I.~G.}\ \bibnamefont {Botella}}, \bibinfo {author} {\bibfnamefont {M.}~\bibnamefont {Lindner}}, \bibinfo {author} {\bibfnamefont {N.}~\bibnamefont {Cartiglia}}, \bibinfo {author} {\bibfnamefont {G.}~\bibnamefont {Pellegrini}}, \bibinfo {author} {\bibfnamefont
  {D.}~\bibnamefont {Bortoletto}}, \bibinfo {author} {\bibfnamefont {D.}~\bibnamefont {Contardo}}, \bibinfo {author} {\bibfnamefont {I.-M.}\ \bibnamefont {Gregor}}, \bibinfo {author} {\bibfnamefont {G.}~\bibnamefont {Kramberger}}, \bibinfo {author} {\bibfnamefont {H.}~\bibnamefont {Pernegger}}, \bibinfo {author} {\bibfnamefont {N.}~\bibnamefont {Harnew}}, \bibinfo {author} {\bibfnamefont {P.}~\bibnamefont {Krizan}}, \bibinfo {author} {\bibfnamefont {I.}~\bibnamefont {Adachi}}, \bibinfo {author} {\bibfnamefont {E.}~\bibnamefont {Nappi}}, \bibinfo {author} {\bibfnamefont {C.}~\bibnamefont {Joram}}, \bibinfo {author} {\bibfnamefont {H.-C.}\ \bibnamefont {Schultz-Coulon}}, \bibinfo {author} {\bibfnamefont {M.}~\bibnamefont {Demarteau}}, \bibinfo {author} {\bibfnamefont {M.}~\bibnamefont {Doser}}, \bibinfo {author} {\bibfnamefont {C.}~\bibnamefont {Braggio}}, \bibinfo {author} {\bibfnamefont {A.}~\bibnamefont {Geraci}}, \bibinfo {author} {\bibfnamefont {P.}~\bibnamefont {Graham}}, \bibinfo {author} {\bibfnamefont
  {A.}~\bibnamefont {Grasselino}}, \bibinfo {author} {\bibfnamefont {J.~M.}\ \bibnamefont {Russell}}, \bibinfo {author} {\bibfnamefont {S.}~\bibnamefont {Withington}}, \bibinfo {author} {\bibfnamefont {R.}~\bibnamefont {Ferrari}}, \bibinfo {author} {\bibfnamefont {R.}~\bibnamefont {Poschl}}, \bibinfo {author} {\bibfnamefont {M.}~\bibnamefont {Aleksa}}, \bibinfo {author} {\bibfnamefont {D.}~\bibnamefont {Barney}}, \bibinfo {author} {\bibfnamefont {F.}~\bibnamefont {Simon}}, \bibinfo {author} {\bibfnamefont {T.~T.~D.}\ \bibnamefont {Fatis}}, \bibinfo {author} {\bibfnamefont {D.}~\bibnamefont {Newbold}}, \bibinfo {author} {\bibfnamefont {F.}~\bibnamefont {Vasey}}, \bibinfo {author} {\bibfnamefont {N.}~\bibnamefont {Neufeld}}, \bibinfo {author} {\bibfnamefont {V.}~\bibnamefont {Re}}, \bibinfo {author} {\bibfnamefont {C.~d.~L.}\ \bibnamefont {Taille}}, \bibinfo {author} {\bibfnamefont {M.}~\bibnamefont {Weber}}, \bibinfo {author} {\bibfnamefont {F.}~\bibnamefont {Hartmann}}, \bibinfo {author} {\bibfnamefont
  {W.}~\bibnamefont {Riegler}}, \bibinfo {author} {\bibfnamefont {C.}~\bibnamefont {Gargiulo}}, \bibinfo {author} {\bibfnamefont {F.}~\bibnamefont {Resnati}}, \bibinfo {author} {\bibfnamefont {H.~T.}\ \bibnamefont {Kate}}, \bibinfo {author} {\bibfnamefont {B.}~\bibnamefont {Verlaat}}, \bibinfo {author} {\bibfnamefont {M.}~\bibnamefont {Vos}}, \bibinfo {author} {\bibfnamefont {J.}~\bibnamefont {Collot}}, \bibinfo {author} {\bibfnamefont {E.}~\bibnamefont {Garutti}}, \bibinfo {author} {\bibfnamefont {R.}~\bibnamefont {Brenner}}, \bibinfo {author} {\bibfnamefont {N.~V.}\ \bibnamefont {Bakel}}, \bibinfo {author} {\bibfnamefont {C.}~\bibnamefont {Gwenlan}}, \bibinfo {author} {\bibfnamefont {J.}~\bibnamefont {Wiener}}, \bibinfo {author} {\bibfnamefont {R.}~\bibnamefont {Appleby}}, \bibinfo {author} {\bibfnamefont {P.}~\bibnamefont {Allport}}, \bibinfo {author} {\bibfnamefont {S.~D.}\ \bibnamefont {Torre}}, \bibinfo {author} {\bibfnamefont {M.}~\bibnamefont {Krammer}}, \bibinfo {author} {\bibfnamefont
  {F.}~\bibnamefont {Sefkow}}, \bibinfo {author} {\bibfnamefont {I.}~\bibnamefont {Shipsey}}, \bibinfo {author} {\bibfnamefont {K.}~\bibnamefont {Jakobs}}, \bibinfo {author} {\bibfnamefont {J.}~\bibnamefont {D'hondt}},\ and\ \bibinfo {author} {\bibfnamefont {L.}~\bibnamefont {Rivkin}},\ }\href {https://doi.org/10.17181/CERN.XDPL.W2EX} {\emph {\bibinfo {title} {{The 2021 ECFA Detector Research and Development Roadmap}}}},\ edited by\ \bibinfo {editor} {\bibnamefont {CERN}}\ (\bibinfo {year} {2021})\BibitemShut {NoStop}%
\bibitem [{\citenamefont {Boyle}\ \emph {et~al.}(2015)\citenamefont {Boyle}, \citenamefont {McEachran}, \citenamefont {Cocks},\ and\ \citenamefont {White}}]{BoyleEtal2015}%
  \BibitemOpen
  \bibfield  {author} {\bibinfo {author} {\bibfnamefont {G.~J.}\ \bibnamefont {Boyle}}, \bibinfo {author} {\bibfnamefont {R.~P.}\ \bibnamefont {McEachran}}, \bibinfo {author} {\bibfnamefont {D.~G.}\ \bibnamefont {Cocks}},\ and\ \bibinfo {author} {\bibfnamefont {R.~D.}\ \bibnamefont {White}},\ }\bibfield  {title} {\bibinfo {title} {{Electron scattering and transport in liquid argon}},\ }\href {https://doi.org/10.1063/1.4917258} {\bibfield  {journal} {\bibinfo  {journal} {The Journal of Chemical Physics}\ }\textbf {\bibinfo {volume} {142}},\ \bibinfo {pages} {154507} (\bibinfo {year} {2015})}\BibitemShut {NoStop}%
\bibitem [{\citenamefont {White}\ \emph {et~al.}(2018)\citenamefont {White}, \citenamefont {Cocks}, \citenamefont {Boyle}, \citenamefont {Casey}, \citenamefont {Garland}, \citenamefont {Konovalov}, \citenamefont {Philippa}, \citenamefont {Stokes}, \citenamefont {de~Urquijo}, \citenamefont {Gonzalez-Magaana}, \citenamefont {McEachran}, \citenamefont {Buckman}, \citenamefont {Brunger}, \citenamefont {Garcia}, \citenamefont {Dujko},\ and\ \citenamefont {Petrovic}}]{WhiteEtal2018}%
  \BibitemOpen
  \bibfield  {author} {\bibinfo {author} {\bibfnamefont {R.~D.}\ \bibnamefont {White}}, \bibinfo {author} {\bibfnamefont {D.}~\bibnamefont {Cocks}}, \bibinfo {author} {\bibfnamefont {G.}~\bibnamefont {Boyle}}, \bibinfo {author} {\bibfnamefont {M.}~\bibnamefont {Casey}}, \bibinfo {author} {\bibfnamefont {N.}~\bibnamefont {Garland}}, \bibinfo {author} {\bibfnamefont {D.}~\bibnamefont {Konovalov}}, \bibinfo {author} {\bibfnamefont {B.}~\bibnamefont {Philippa}}, \bibinfo {author} {\bibfnamefont {P.}~\bibnamefont {Stokes}}, \bibinfo {author} {\bibfnamefont {J.}~\bibnamefont {de~Urquijo}}, \bibinfo {author} {\bibfnamefont {O.}~\bibnamefont {Gonzalez-Magaana}}, \bibinfo {author} {\bibfnamefont {R.~P.}\ \bibnamefont {McEachran}}, \bibinfo {author} {\bibfnamefont {S.~J.}\ \bibnamefont {Buckman}}, \bibinfo {author} {\bibfnamefont {M.~J.}\ \bibnamefont {Brunger}}, \bibinfo {author} {\bibfnamefont {G.}~\bibnamefont {Garcia}}, \bibinfo {author} {\bibfnamefont {S.}~\bibnamefont {Dujko}},\ and\ \bibinfo {author}
  {\bibfnamefont {Z.~L.}\ \bibnamefont {Petrovic}},\ }\bibfield  {title} {\bibinfo {title} {Electron transport in biomolecular gaseous and liquid systems: theory, experiment and self-consistent cross-sections},\ }\href {https://doi.org/10.1088/1361-6595/aabdd7} {\bibfield  {journal} {\bibinfo  {journal} {Plasma Sources Science and Technology}\ }\textbf {\bibinfo {volume} {27}},\ \bibinfo {pages} {053001} (\bibinfo {year} {2018})}\BibitemShut {NoStop}%
\bibitem [{\citenamefont {Boyle}\ \emph {et~al.}(2016)\citenamefont {Boyle}, \citenamefont {McEachran}, \citenamefont {Cocks}, \citenamefont {Brunger}, \citenamefont {Buckman}, \citenamefont {Dujko},\ and\ \citenamefont {White}}]{BoyleEtal2016}%
  \BibitemOpen
  \bibfield  {author} {\bibinfo {author} {\bibfnamefont {G.~J.}\ \bibnamefont {Boyle}}, \bibinfo {author} {\bibfnamefont {R.~P.}\ \bibnamefont {McEachran}}, \bibinfo {author} {\bibfnamefont {D.~G.}\ \bibnamefont {Cocks}}, \bibinfo {author} {\bibfnamefont {M.~J.}\ \bibnamefont {Brunger}}, \bibinfo {author} {\bibfnamefont {S.~J.}\ \bibnamefont {Buckman}}, \bibinfo {author} {\bibfnamefont {S.}~\bibnamefont {Dujko}},\ and\ \bibinfo {author} {\bibfnamefont {R.~D.}\ \bibnamefont {White}},\ }\bibfield  {title} {\bibinfo {title} {Ab initio electron scattering cross-sections and transport in liquid xenon},\ }\href {https://doi.org/10.1088/0022-3727/49/35/355201} {\bibfield  {journal} {\bibinfo  {journal} {Journal of Physics D: Applied Physics}\ }\textbf {\bibinfo {volume} {49}},\ \bibinfo {pages} {355201} (\bibinfo {year} {2016})}\BibitemShut {NoStop}%
\bibitem [{\citenamefont {Lekner}(1967)}]{Lekner1967}%
  \BibitemOpen
  \bibfield  {author} {\bibinfo {author} {\bibfnamefont {J.}~\bibnamefont {Lekner}},\ }\bibfield  {title} {\bibinfo {title} {Motion of electrons in liquid argon},\ }\href {https://doi.org/10.1103/PhysRev.158.130} {\bibfield  {journal} {\bibinfo  {journal} {Phys. Rev.}\ }\textbf {\bibinfo {volume} {158}},\ \bibinfo {pages} {130} (\bibinfo {year} {1967})}\BibitemShut {NoStop}%
\bibitem [{\citenamefont {White}\ and\ \citenamefont {Robson}(2011)}]{WhitRobs11}%
  \BibitemOpen
  \bibfield  {author} {\bibinfo {author} {\bibfnamefont {R.~D.}\ \bibnamefont {White}}\ and\ \bibinfo {author} {\bibfnamefont {R.~E.}\ \bibnamefont {Robson}},\ }\bibfield  {title} {\bibinfo {title} {Multiterm solution of a generalized {B}oltzmann kinetic equation for electron and positron transport in structured and soft condensed matter},\ }\href@noop {} {\bibfield  {journal} {\bibinfo  {journal} {Physical Review E}\ }\textbf {\bibinfo {volume} {84}},\ \bibinfo {pages} {031125} (\bibinfo {year} {2011})}\BibitemShut {NoStop}%
\bibitem [{\citenamefont {Boyle}\ \emph {et~al.}(2017)\citenamefont {Boyle}, \citenamefont {Tattersall}, \citenamefont {Cocks}, \citenamefont {McEachran},\ and\ \citenamefont {White}}]{Boyl17}%
  \BibitemOpen
  \bibfield  {author} {\bibinfo {author} {\bibfnamefont {G.~J.}\ \bibnamefont {Boyle}}, \bibinfo {author} {\bibfnamefont {W.~J.}\ \bibnamefont {Tattersall}}, \bibinfo {author} {\bibfnamefont {D.~G.}\ \bibnamefont {Cocks}}, \bibinfo {author} {\bibfnamefont {R.~P.}\ \bibnamefont {McEachran}},\ and\ \bibinfo {author} {\bibfnamefont {R.~D.}\ \bibnamefont {White}},\ }\bibfield  {title} {\bibinfo {title} {A multi-term solution of the space–time {B}oltzmann equation for electrons in gases and liquids},\ }\href {https://doi.org/10.1088/1361-6595/aa51ef} {\bibfield  {journal} {\bibinfo  {journal} {Plasma Sources Science and Technology}\ }\textbf {\bibinfo {volume} {26}},\ \bibinfo {pages} {024007} (\bibinfo {year} {2017})}\BibitemShut {NoStop}%
\bibitem [{\citenamefont {{Blanc, A.}}(1908)}]{Blanc1908}%
  \BibitemOpen
  \bibfield  {author} {\bibinfo {author} {\bibnamefont {{Blanc, A.}}},\ }\bibfield  {title} {\bibinfo {title} {Recherches sur les mobilit\'es des ions dans les gaz},\ }\href {https://doi.org/10.1051/jphystap:019080070082501} {\bibfield  {journal} {\bibinfo  {journal} {J. Phys. Theor. Appl.}\ }\textbf {\bibinfo {volume} {7}},\ \bibinfo {pages} {825} (\bibinfo {year} {1908})}\BibitemShut {NoStop}%
\bibitem [{\citenamefont {Hansen}\ and\ \citenamefont {McDonald}(1976)}]{HansenMcDonald1973}%
  \BibitemOpen
  \bibfield  {author} {\bibinfo {author} {\bibfnamefont {J.-P.}\ \bibnamefont {Hansen}}\ and\ \bibinfo {author} {\bibfnamefont {I.~R.}\ \bibnamefont {McDonald}},\ }\href@noop {} {\emph {\bibinfo {title} {Theory of simple liquids}}}\ (\bibinfo  {publisher} {Academic press},\ \bibinfo {year} {1976})\BibitemShut {NoStop}%
\bibitem [{\citenamefont {Ashcroft}\ and\ \citenamefont {Langreth}(1967)}]{AshcroftLangreth1967}%
  \BibitemOpen
  \bibfield  {author} {\bibinfo {author} {\bibfnamefont {N.}~\bibnamefont {Ashcroft}}\ and\ \bibinfo {author} {\bibfnamefont {D.~C.}\ \bibnamefont {Langreth}},\ }\bibfield  {title} {\bibinfo {title} {Structure of binary liquid mixtures. {I}},\ }\href@noop {} {\bibfield  {journal} {\bibinfo  {journal} {Physical Review}\ }\textbf {\bibinfo {volume} {156}},\ \bibinfo {pages} {685} (\bibinfo {year} {1967})}\BibitemShut {NoStop}%
\bibitem [{\citenamefont {van Gunsteren}\ and\ \citenamefont {Berendsen}(1990)}]{GunsterenBeredsen90}%
  \BibitemOpen
  \bibfield  {author} {\bibinfo {author} {\bibfnamefont {W.~F.}\ \bibnamefont {van Gunsteren}}\ and\ \bibinfo {author} {\bibfnamefont {H.~J.~C.}\ \bibnamefont {Berendsen}},\ }\bibfield  {title} {\bibinfo {title} {Computer simulation of molecular dynamics: Methodology, applications, and perspectives in chemistry},\ }\href {https://doi.org/https://doi.org/10.1002/anie.199009921} {\bibfield  {journal} {\bibinfo  {journal} {Angewandte Chemie International Edition in English}\ }\textbf {\bibinfo {volume} {29}},\ \bibinfo {pages} {992} (\bibinfo {year} {1990})}\BibitemShut {NoStop}%
\bibitem [{\citenamefont {Lindg{\aa}rd}(1994)}]{Lindgard1994}%
  \BibitemOpen
  \bibfield  {author} {\bibinfo {author} {\bibfnamefont {P.~A.}\ \bibnamefont {Lindg{\aa}rd}},\ }\bibfield  {title} {\bibinfo {title} {Computer simulation of the structure factor},\ }in\ \href@noop {} {\emph {\bibinfo {booktitle} {Computer Simulation Studies in Condensed-Matter Physics VII}}},\ \bibinfo {editor} {edited by\ \bibinfo {editor} {\bibfnamefont {D.~P.}\ \bibnamefont {Landau}}, \bibinfo {editor} {\bibfnamefont {K.~K.}\ \bibnamefont {Mon}},\ and\ \bibinfo {editor} {\bibfnamefont {H.-B.}\ \bibnamefont {Sch{\"u}ttler}}}\ (\bibinfo  {publisher} {Springer Berlin Heidelberg},\ \bibinfo {address} {Berlin, Heidelberg},\ \bibinfo {year} {1994})\ pp.\ \bibinfo {pages} {69--82}\BibitemShut {NoStop}%
\bibitem [{\citenamefont {Wedberg}\ \emph {et~al.}(2011)\citenamefont {Wedberg}, \citenamefont {O’Connell}, \citenamefont {Peters},\ and\ \citenamefont {Abildskov}}]{WedbergEtal2011}%
  \BibitemOpen
  \bibfield  {author} {\bibinfo {author} {\bibfnamefont {R.}~\bibnamefont {Wedberg}}, \bibinfo {author} {\bibfnamefont {J.~P.}\ \bibnamefont {O’Connell}}, \bibinfo {author} {\bibfnamefont {G.~H.}\ \bibnamefont {Peters}},\ and\ \bibinfo {author} {\bibfnamefont {J.}~\bibnamefont {Abildskov}},\ }\bibfield  {title} {\bibinfo {title} {{Pair correlation function integrals: Computation and use}},\ }\href {https://doi.org/10.1063/1.3626799} {\bibfield  {journal} {\bibinfo  {journal} {The Journal of Chemical Physics}\ }\textbf {\bibinfo {volume} {135}},\ \bibinfo {pages} {084113} (\bibinfo {year} {2011})}\BibitemShut {NoStop}%
\bibitem [{\citenamefont {Percus}\ and\ \citenamefont {Yevick}(1958)}]{PercusYevick1958}%
  \BibitemOpen
  \bibfield  {author} {\bibinfo {author} {\bibfnamefont {J.~K.}\ \bibnamefont {Percus}}\ and\ \bibinfo {author} {\bibfnamefont {G.~J.}\ \bibnamefont {Yevick}},\ }\bibfield  {title} {\bibinfo {title} {Analysis of classical statistical mechanics by means of collective coordinates},\ }\href {https://doi.org/10.1103/PhysRev.110.1} {\bibfield  {journal} {\bibinfo  {journal} {Phys. Rev.}\ }\textbf {\bibinfo {volume} {110}},\ \bibinfo {pages} {1} (\bibinfo {year} {1958})}\BibitemShut {NoStop}%
\bibitem [{\citenamefont {Hoshino}(1983)}]{Hoshino1983}%
  \BibitemOpen
  \bibfield  {author} {\bibinfo {author} {\bibfnamefont {K.}~\bibnamefont {Hoshino}},\ }\bibfield  {title} {\bibinfo {title} {Structure of multi-component hard-sphere mixtures- application to the liquid li-pb alloy},\ }\href {https://doi.org/10.1088/0305-4608/13/10/010} {\bibfield  {journal} {\bibinfo  {journal} {Journal of Physics F: Metal Physics}\ }\textbf {\bibinfo {volume} {13}},\ \bibinfo {pages} {1981} (\bibinfo {year} {1983})}\BibitemShut {NoStop}%
\bibitem [{\citenamefont {Kumar}\ \emph {et~al.}(1980)\citenamefont {Kumar}, \citenamefont {Skullerud},\ and\ \citenamefont {Robson}}]{KumaSkulRobs80}%
  \BibitemOpen
  \bibfield  {author} {\bibinfo {author} {\bibfnamefont {K.}~\bibnamefont {Kumar}}, \bibinfo {author} {\bibfnamefont {H.}~\bibnamefont {Skullerud}},\ and\ \bibinfo {author} {\bibfnamefont {R.~E.}\ \bibnamefont {Robson}},\ }\bibfield  {title} {\bibinfo {title} {Kinetic theory of charged particle swarms in neutral gases},\ }\href@noop {} {\bibfield  {journal} {\bibinfo  {journal} {Australian Journal of Physics}\ }\textbf {\bibinfo {volume} {33}},\ \bibinfo {pages} {343} (\bibinfo {year} {1980})}\BibitemShut {NoStop}%
\bibitem [{\citenamefont {Abramowitz}\ and\ \citenamefont {Stegun}(1972)}]{AbramowitzStegun1972}%
  \BibitemOpen
  \bibfield  {author} {\bibinfo {author} {\bibfnamefont {M.}~\bibnamefont {Abramowitz}}\ and\ \bibinfo {author} {\bibfnamefont {I.}~\bibnamefont {Stegun}},\ }\href@noop {} {\emph {\bibinfo {title} {Handbook of mathematical functions}}}\ (\bibinfo  {publisher} {Dover Publications Inc.},\ \bibinfo {year} {1972})\BibitemShut {NoStop}%
\bibitem [{\citenamefont {Robson}\ \emph {et~al.}(2017)\citenamefont {Robson}, \citenamefont {White},\ and\ \citenamefont {Hildebrandt}}]{RobsWhitHild17}%
  \BibitemOpen
  \bibfield  {author} {\bibinfo {author} {\bibfnamefont {R.~E.}\ \bibnamefont {Robson}}, \bibinfo {author} {\bibfnamefont {R.~D.}\ \bibnamefont {White}},\ and\ \bibinfo {author} {\bibfnamefont {M.}~\bibnamefont {Hildebrandt}},\ }\href@noop {} {\emph {\bibinfo {title} {Fundamentals of Charged Particle Transport in Gases and Condensed Matter}}}\ (\bibinfo  {publisher} {CRC Press},\ \bibinfo {year} {2017})\BibitemShut {NoStop}%
\bibitem [{\citenamefont {Fitzpatrick}(2015)}]{Fitzpatrick2015}%
  \BibitemOpen
  \bibfield  {author} {\bibinfo {author} {\bibfnamefont {R.}~\bibnamefont {Fitzpatrick}},\ }\href {https://doi.org/10.1142/9645} {\emph {\bibinfo {title} {Quantum Mechanics}}}\ (\bibinfo  {publisher} {World Scientific},\ \bibinfo {year} {2015})\BibitemShut {NoStop}%
\bibitem [{\citenamefont {Tattersall}\ \emph {et~al.}(2015)\citenamefont {Tattersall}, \citenamefont {Cocks}, \citenamefont {Boyle}, \citenamefont {Buckman},\ and\ \citenamefont {White}}]{TattersallEtal2015}%
  \BibitemOpen
  \bibfield  {author} {\bibinfo {author} {\bibfnamefont {W.~J.}\ \bibnamefont {Tattersall}}, \bibinfo {author} {\bibfnamefont {D.~G.}\ \bibnamefont {Cocks}}, \bibinfo {author} {\bibfnamefont {G.~J.}\ \bibnamefont {Boyle}}, \bibinfo {author} {\bibfnamefont {S.~J.}\ \bibnamefont {Buckman}},\ and\ \bibinfo {author} {\bibfnamefont {R.~D.}\ \bibnamefont {White}},\ }\bibfield  {title} {\bibinfo {title} {Monte carlo study of coherent scattering effects of low-energy charged particle transport in {P}ercus-{Y}evick liquids},\ }\href {https://doi.org/10.1103/PhysRevE.91.043304} {\bibfield  {journal} {\bibinfo  {journal} {Phys. Rev. E}\ }\textbf {\bibinfo {volume} {91}},\ \bibinfo {pages} {043304} (\bibinfo {year} {2015})}\BibitemShut {NoStop}%
\bibitem [{\citenamefont {White}\ and\ \citenamefont {Robson}(2009)}]{WhiteRobs09}%
  \BibitemOpen
  \bibfield  {author} {\bibinfo {author} {\bibfnamefont {R.~D.}\ \bibnamefont {White}}\ and\ \bibinfo {author} {\bibfnamefont {R.~E.}\ \bibnamefont {Robson}},\ }\bibfield  {title} {\bibinfo {title} {Positron kinetics in soft condensed matter},\ }\href {https://doi.org/10.1103/PhysRevLett.102.230602} {\bibfield  {journal} {\bibinfo  {journal} {Phys. Rev. Lett.}\ }\textbf {\bibinfo {volume} {102}},\ \bibinfo {pages} {230602} (\bibinfo {year} {2009})}\BibitemShut {NoStop}%
\bibitem [{\citenamefont {White}\ \emph {et~al.}(2003)\citenamefont {White}, \citenamefont {Robson}, \citenamefont {Schmidt},\ and\ \citenamefont {Morrison}}]{WhitEtal03}%
  \BibitemOpen
  \bibfield  {author} {\bibinfo {author} {\bibfnamefont {R.~D.}\ \bibnamefont {White}}, \bibinfo {author} {\bibfnamefont {R.~E.}\ \bibnamefont {Robson}}, \bibinfo {author} {\bibfnamefont {B.}~\bibnamefont {Schmidt}},\ and\ \bibinfo {author} {\bibfnamefont {M.~A.}\ \bibnamefont {Morrison}},\ }\bibfield  {title} {\bibinfo {title} {Is the classical two-term approximation of electron kinetic theory satisfactory for swarms and plasmas?},\ }\href@noop {} {\bibfield  {journal} {\bibinfo  {journal} {Journal of Physics D: applied physics}\ }\textbf {\bibinfo {volume} {36}},\ \bibinfo {pages} {3125} (\bibinfo {year} {2003})}\BibitemShut {NoStop}%
\bibitem [{\citenamefont {Petrovi{\'c}}\ \emph {et~al.}(2009)\citenamefont {Petrovi{\'c}}, \citenamefont {Dujko}, \citenamefont {Mari{\'c}}, \citenamefont {Malovi{\'c}}, \citenamefont {Nikitovi{\'c}}, \citenamefont {Sasi{\'c}}, \citenamefont {Jovanovi{\'c}}, \citenamefont {Stojanovi{\'c}},\ and\ \citenamefont {Radmilovi{\'c}-Radjenovi{\'c}}}]{PetrEtal09}%
  \BibitemOpen
  \bibfield  {author} {\bibinfo {author} {\bibfnamefont {Z.~L.}\ \bibnamefont {Petrovi{\'c}}}, \bibinfo {author} {\bibfnamefont {S.}~\bibnamefont {Dujko}}, \bibinfo {author} {\bibfnamefont {D.}~\bibnamefont {Mari{\'c}}}, \bibinfo {author} {\bibfnamefont {G.}~\bibnamefont {Malovi{\'c}}}, \bibinfo {author} {\bibfnamefont {Z.}~\bibnamefont {Nikitovi{\'c}}}, \bibinfo {author} {\bibfnamefont {O.}~\bibnamefont {Sasi{\'c}}}, \bibinfo {author} {\bibfnamefont {J.}~\bibnamefont {Jovanovi{\'c}}}, \bibinfo {author} {\bibfnamefont {V.}~\bibnamefont {Stojanovi{\'c}}},\ and\ \bibinfo {author} {\bibfnamefont {M.}~\bibnamefont {Radmilovi{\'c}-Radjenovi{\'c}}},\ }\bibfield  {title} {\bibinfo {title} {Measurement and interpretation of swarm parameters and their application in plasma modelling},\ }\href@noop {} {\bibfield  {journal} {\bibinfo  {journal} {Journal of Physics D: Applied Physics}\ }\textbf {\bibinfo {volume} {42}},\ \bibinfo {pages} {194002} (\bibinfo {year} {2009})}\BibitemShut {NoStop}%
\bibitem [{\citenamefont {Šašić}\ \emph {et~al.}(2005)\citenamefont {Šašić}, \citenamefont {Jovanović}, \citenamefont {Petrović}, \citenamefont {de~Urquijo}, \citenamefont {Castrejón-Pita}, \citenamefont {Hernández-Ávila},\ and\ \citenamefont {Basurto}}]{SasicEtal2005}%
  \BibitemOpen
  \bibfield  {author} {\bibinfo {author} {\bibfnamefont {O.}~\bibnamefont {Šašić}}, \bibinfo {author} {\bibfnamefont {J.}~\bibnamefont {Jovanović}}, \bibinfo {author} {\bibfnamefont {Z.~L.}\ \bibnamefont {Petrović}}, \bibinfo {author} {\bibfnamefont {J.}~\bibnamefont {de~Urquijo}}, \bibinfo {author} {\bibfnamefont {J.~R.}\ \bibnamefont {Castrejón-Pita}}, \bibinfo {author} {\bibfnamefont {J.~L.}\ \bibnamefont {Hernández-Ávila}},\ and\ \bibinfo {author} {\bibfnamefont {E.}~\bibnamefont {Basurto}},\ }\bibfield  {title} {\bibinfo {title} {Electron drift velocities in mixtures of helium and xenon and experimental verification of corrections to {Blanc}'s law},\ }\href {https://doi.org/10.1103/PhysRevE.71.046408} {\bibfield  {journal} {\bibinfo  {journal} {Physical Review E}\ }\textbf {\bibinfo {volume} {71}},\ \bibinfo {pages} {046408} (\bibinfo {year} {2005})}\BibitemShut {NoStop}%
\bibitem [{\citenamefont {Wang}\ and\ \citenamefont {Van~Brunt}(1997)}]{WangBrunt1997}%
  \BibitemOpen
  \bibfield  {author} {\bibinfo {author} {\bibfnamefont {Y.}~\bibnamefont {Wang}}\ and\ \bibinfo {author} {\bibfnamefont {R.~J.}\ \bibnamefont {Van~Brunt}},\ }\bibfield  {title} {\bibinfo {title} {Calculation of electron transport in {Ar}/{N$_2$} and {He}/{Kr} gas mixtures—implications for validity of the {Blanc}’s law method},\ }\href {https://doi.org/10.1063/1.872563} {\bibfield  {journal} {\bibinfo  {journal} {Physics of Plasmas}\ }\textbf {\bibinfo {volume} {4}},\ \bibinfo {pages} {551} (\bibinfo {year} {1997})}\BibitemShut {NoStop}%
\bibitem [{\citenamefont {Garland}\ \emph {et~al.}(2018)\citenamefont {Garland}, \citenamefont {Boyle}, \citenamefont {Cocks},\ and\ \citenamefont {White}}]{GarlandEtal2018}%
  \BibitemOpen
  \bibfield  {author} {\bibinfo {author} {\bibfnamefont {N.~A.}\ \bibnamefont {Garland}}, \bibinfo {author} {\bibfnamefont {G.~J.}\ \bibnamefont {Boyle}}, \bibinfo {author} {\bibfnamefont {D.~G.}\ \bibnamefont {Cocks}},\ and\ \bibinfo {author} {\bibfnamefont {R.~D.}\ \bibnamefont {White}},\ }\bibfield  {title} {\bibinfo {title} {Approximating the nonlinear density dependence of electron transport coefficients and scattering rates across the gas-liquid interface},\ }\href {https://doi.org/10.1088/1361-6595/aaaa0c} {\bibfield  {journal} {\bibinfo  {journal} {Plasma Sources Science and Technology}\ }\textbf {\bibinfo {volume} {27}},\ \bibinfo {pages} {024002} (\bibinfo {year} {2018})}\BibitemShut {NoStop}%
\bibitem [{\citenamefont {Borghesani}\ \emph {et~al.}(1993)\citenamefont {Borghesani}, \citenamefont {Cerdonio}, \citenamefont {Conti}, \citenamefont {Onofrio}, \citenamefont {Bressi},\ and\ \citenamefont {Carugno}}]{Borghesani1993}%
  \BibitemOpen
  \bibfield  {author} {\bibinfo {author} {\bibfnamefont {A.}~\bibnamefont {Borghesani}}, \bibinfo {author} {\bibfnamefont {S.}~\bibnamefont {Cerdonio}}, \bibinfo {author} {\bibfnamefont {E.}~\bibnamefont {Conti}}, \bibinfo {author} {\bibfnamefont {R.}~\bibnamefont {Onofrio}}, \bibinfo {author} {\bibfnamefont {G.}~\bibnamefont {Bressi}},\ and\ \bibinfo {author} {\bibfnamefont {G.}~\bibnamefont {Carugno}},\ }\bibfield  {title} {\bibinfo {title} {Excess electron mobility in liquid {Xe}—{Ar} mixtures},\ }\href {https://doi.org/https://doi.org/10.1016/0375-9601(93)90869-2} {\bibfield  {journal} {\bibinfo  {journal} {Physics Letters A}\ }\textbf {\bibinfo {volume} {178}},\ \bibinfo {pages} {407} (\bibinfo {year} {1993})}\BibitemShut {NoStop}%
\bibitem [{\citenamefont {Borghesani}\ \emph {et~al.}(1997)\citenamefont {Borghesani}, \citenamefont {Iannuzzi},\ and\ \citenamefont {Carugno}}]{Borghesani1997}%
  \BibitemOpen
  \bibfield  {author} {\bibinfo {author} {\bibfnamefont {A.~F.}\ \bibnamefont {Borghesani}}, \bibinfo {author} {\bibfnamefont {D.}~\bibnamefont {Iannuzzi}},\ and\ \bibinfo {author} {\bibfnamefont {G.}~\bibnamefont {Carugno}},\ }\bibfield  {title} {\bibinfo {title} {Excess electron mobility in liquid {Ar}-{Kr} and {Ar}-{Xe} mixtures},\ }\href {https://doi.org/10.1088/0953-8984/9/24/006} {\bibfield  {journal} {\bibinfo  {journal} {Journal of Physics: Condensed Matter}\ }\textbf {\bibinfo {volume} {9}},\ \bibinfo {pages} {5057} (\bibinfo {year} {1997})}\BibitemShut {NoStop}%
\end{thebibliography}%

\end{document}